# Best practices for machine learning in antibody discovery and development

Leonard Wossnig[1,2,†], Norbert Furtmann[3], Andrew Buchanan[4], Sandeep Kumar[5], and Victor Greiff[6]

[1] LabGenius Ltd., The Biscuit Factory, 100 Drummond Road, London, SE16 4DG London, United Kingdom; Orcid: 0000-0002-0861-9540
[2] Department of Computer Science, University College London, 66-72 Gower St, WC1E 6EA London, United Kingdom
[3] R&D Large Molecules Research Platform, Sanofi Deutschland GmbH, Industriepark Höchst, Frankfurt Am Main, Germany; Orcid: 0009-0006-8226-7586
[4] Biologics Engineering, R&D, AstraZeneca, Cambridge, CB2 0AA, United Kingdom; Orcid: 0000-0002-5191-7682
[5] Computational Protein Design and Modeling Group, Computational Science, Moderna Therapeutics, 200 Technology Square, Cambridge, MA 02139, United States of America; Orcid: 0000-0003-2840-6398
[6] Department of Immunology and Oslo University Hospital, University of Oslo, Oslo, Norway; Orcid: 0000-0003-2622-5032

† Corresponding author

**Abstract:**
Over the past 40 years, the discovery and development of therapeutic antibodies to treat disease has become common practice. However, as therapeutic antibody constructs are becoming more sophisticated (e.g., multi-specifics), conventional approaches to optimisation are increasingly inefficient. Machine learning (ML) promises to open up an *in silico* route to antibody discovery and help accelerate the development of drug products using a reduced number of experiments and hence cost.

Over the past few years, we have observed rapid developments in the field of ML-guided antibody discovery and development (D&D). However, many of the results are difficult to compare or hard to assess for utility by other experts in the field due to the high diversity in the datasets and evaluation techniques and metrics that are across industry and academia. This limitation of the literature curtails the broad adoption of ML across the industry and slows down overall progress in the field, highlighting the need to develop standards and guidelines that may help improve the reproducibility of ML models across different research groups.

To address these challenges, we set out in this perspective to critically review current practices, explain common pitfalls, and clearly define a set of method development and evaluation guidelines that can be applied to different types of ML-based techniques for therapeutic antibody D&D. Specifically, we address in an end-to-end analysis, challenges associated with all aspects of the ML process and recommend a set of best practices for each stage.

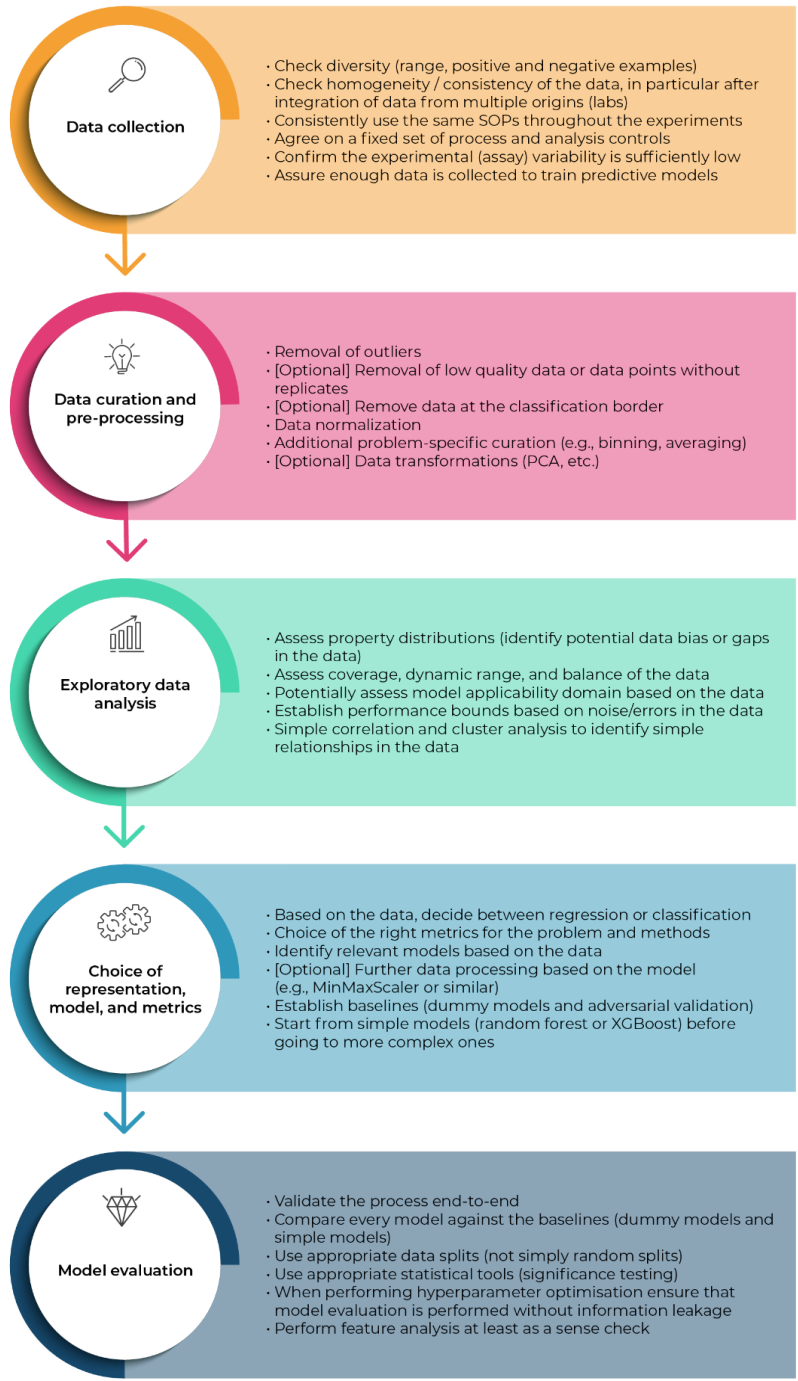

**Figure 1:** Overview of the entire machine learning (ML) process for antibody R&D from data collection to model evaluation.

# 1. Introduction

The development of antibody-based drugs has revolutionised the field of medicine, providing effective treatments for a wide range of diseases [1,2]. However, although the drug discovery process has proven effective, it is complex, expensive, time-consuming, and has room for improvement [3–7]. Recent advances in ML have the potential to accelerate and optimise this process by enabling the identification of better biotherapeutic drug candidates more rapidly, and thereby reducing the cost and timelines for antibody drug discovery [8–10].

The rapidly evolving landscape of ML-guided therapeutic antibody research and development (R&D) holds immense potential for the biopharmaceutical industry. Although initial success stories have emerged, its 'real world' impact is thus far relatively minimal [11]. To unlock the full potential and demonstrate significant impact in commercial drug discovery and development settings, it is essential to establish standardised guidelines and best practices for applications of ML at every step of biologic drug discovery and development projects. These include *in-silico* design of antibody candidate drugs; computational identification or design of high affinity specific function relevant epitopes; accurate prediction of biophysical attributes; screening for formulations; and development of digital twins of manufacturing processes. [12,13].

For next-generation multispecific antibodies, these computational approaches are essential for assessing the large design spaces and identifying candidates with optimal properties that can then progress towards clinical studies. Improving the probability of success in clinical trials generally has the highest positive impact on the costs and timelines for an individual program [14]. Nevertheless, a reduction of time and costs in the preclinical stages can still enable significant overall savings due to the high volume of early stage programs that are executed in parallel throughout the industry, assuming the resulting biotherapeutics are a similar quality or better than those conventionally discovered [15]. Recent years have, therefore, witnessed a surge in the application of ML-based models at each stage of the drug discovery and development cycle. However, little to no attention is being paid to benchmark general applicability as well as benchmarking of these models, leading to difficulties reproducing models and results.

This review article aims to address the need for establishing benchmarks and best practices for the use of ML in the biopharmaceutical industry. We will critically examine current methodologies, identifying common pitfalls and providing recommendations for ML-based approaches to therapeutic antibody R&D akin to other reviews for chemistry or broader ML research [16–19]. Unlike other reviews, (see *e.g.* [20–24]) we are less concerned with what type of model to use in which context and the presentation of different ML methods. Instead, we focus heavily on data aspects and model validation. This focus stems from our practical drug discovery expertise, where the latter are crucial factors that have been commonly neglected or undervalued in other reviews of this emerging field.

By offering clear evaluation guidelines based on literature and experience from practitioners from both small biotech and large pharma, we aim to bridge the gap between theoretical advancements and real-world applications. The focus is on therapeutic antibody engineering, and aims to foster consistency across academic and industry research groups. We believe that this, in turn, will enable ML to contribute more meaningfully to the R&D of novel antibody therapeutics and ultimately benefit the patients.

While the main focus of this review is on the creation of high quality data sets and ML models, we need to touch on the most fundamental problem that needs to be addressed first when working on any drug discovery program: The underlying objective. This topic could be summarised under the broad term of 'predictive validity', *i.e.*, are we increasing confidence in the project's validation for likely product development and clinical success. In other words, are we increasing confidence that we have a developable antibody which is safe, binds only the intended target and produces the desired therapeutic effect via the desired mechanism of action (MoA) among the correct patient group [25]? Only once this is clearly answered, should a team move on to building data sets and ML models to support a given program.

**2. Before you start: Choose the appropriate strategy and the experimental set-up**

Before discussing the intricacies of a ML process, any antibody discovery program needs to have a clear experimental strategy aligned to the candidate drug target profile (CDTP) [26]. The CDTP should include targets for both the desired pharmacology and the appropriate developability package (Fig. 2 A). This specification sheet will inform both the assay cascade for lead generation, optimisation and final candidate selection (Fig. 2 B).

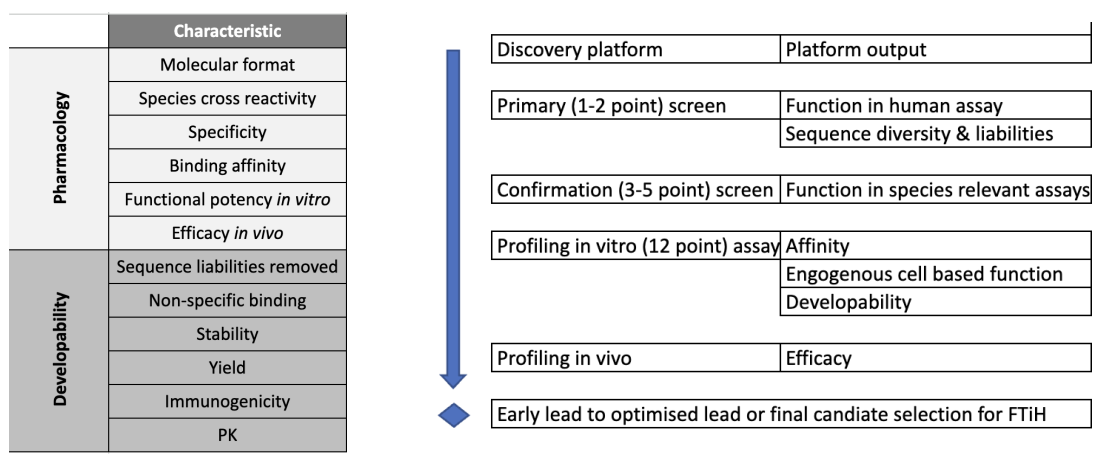

**Figure 2:** (A) An illustrative candidate drug target profile (CDTP) for any biologic modality covering key attributes of both pharmacology and developability. The targets should be set appropriately for the relevant milestones during R&D

towards candidate selection. (B) An outline to illustrate an assay cascade for lead selection during R&D. Successful completion may be achieved following one or a number of iterations of the cascade dependent upon aspects of the platform and biology of the target.

In drug discovery, building confidence in both the target and the candidate molecule's pharmacology and developability are key to progress from target selection to candidate drug selection prior to clinical and CMC investment. CMC science includes cell line development, formulation and drug delivery, device development, analytical capabilities and clinical manufacture and supply for First Time in Human (FTiH) clinical studies to launch. To maximise the probability of success establishing assays that relate to species cross reactivity, desired MoA and aspects of developability is crucial. Focusing on function rather than affinity or other simplistic properties enables the identification of the rarer functional biologics, especially in the context of complex targets, agonists, multispecific or multivalent biologics [27–29]. To further stress this point, many screening or optimisation campaigns still primarily focus on affinity maturation, even though more predictive assays (*e.g.* functional, reporter assays), which could deliver a similar throughput, are available today. In most instances these binding optimisation campaigns have limited correlation with the desired function, even for simple biologies such as agonists (see Fig. 3).

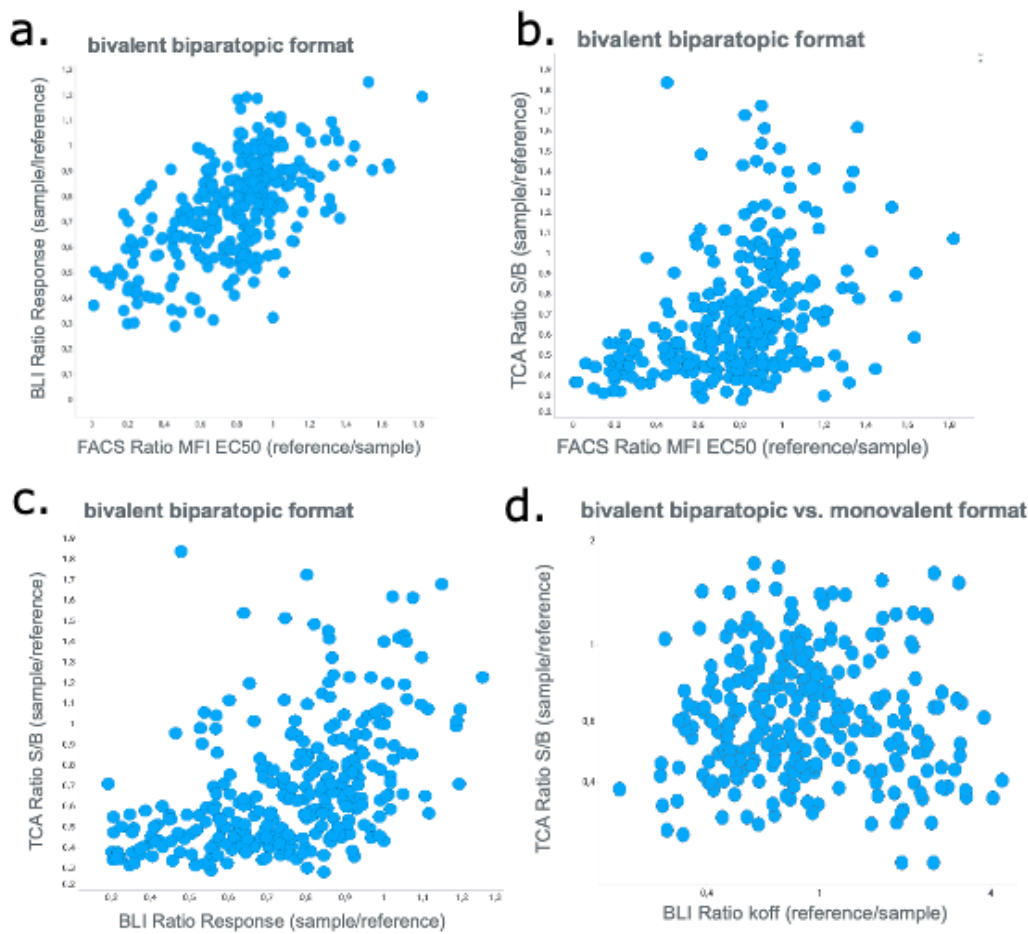

**Figure 3.** The figure illustrates correlations between binding and activity behaviours assessed using different technologies for VHH domains in both mono- and bivalent (biparatopic) formats against a specific therapeutic target that requires agonistic activity. VHH-target binding was evaluated through biolayer interferometry (BLI), cell binding through FACS, and activity within a T cell activation assay. In this context, agonistic activity is achievable only by combining two binding domains against the same target within a biparatopic format.

In Figure 3a, binding against the immobilised target measured via BLI (response) correlates with cell binding assessed via FACS (EC50) for compounds in the biparatopic format. However, Figures 3b and 3c demonstrate that binding to the immobilised target (BLI response) and binding to cells (FACS EC50) poorly correlate with activity in the T cell activation assay. Figure 3d illustrates that the binding (BLI off-rates) of monovalent building blocks does not correlate at all with the activation behaviour of the same variants in the biparatopic format in the T-cell activation assay.

Appropriate functional assays can be established in high-throughput (HT) screening mode or lower throughput profiling mode with increasing translational relevance. Examples of HT functional screening assays are biochemical ligand/receptor HTRF® assays or cell-based reporter systems. Progressing from engineered biochemical and cell-based *in vitro* assays to more disease relevant cell-based assays with endogenous target biology or patient derived systems increases translational relevance but can reduce throughput and precision, and increase variation.

Designing a project specific assay cascade entails balancing assay feasibility, throughput, robustness, data quality and translational relevance. For each project, decisions are required for what data can best triage libraries and molecules through lead generation and optimisation phases. The integration of wet-lab automation with the goal of generating structured, consistent, machine-readable data will enhance data generation efficiency and accuracy [30,31]. Data accuracy, data amount, and time to complete a full loop from prediction to experimental validation all need careful consideration for an optimal predictive performance and hence project outcome [32].

An example assay cascade with approximate numbers and endpoints is outlined in Table 1. The triaging of assays, throughput, and phasing depends upon individual project requirements and biological feasibility. Key considerations in successfully implementing this revolve around technical and cultural ways of working. This includes aspects of: standardisation of labware and vendors; having skilled automation engineers to code and optimise the wet-lab scientist methods; use of barcoding and tracking samples from sequence through expression and assay systems; the use of appropriate controls across users, protocols and projects; standardised parsing analysis to route data back to the end user. Staff's real-time access to integrated and curated data will also facilitate better and more timely decision-making. Other aspects of data generation, capture and analysis to enable ML are summarised in Box 1.

One aspect that is important for ML purposes is the number of dilution points in concentration-response curves. In lead generation or early lead optimization a single point or a small number (*e.g.* 3) of concentrations are often chosen to enable screening of a larger number of candidates in a given run. For later stages, a higher number of points are often chosen in order to obtain more accurate EC50 or IC50, which typically comes at the cost of a lower number of tested designs. From a ML perspective, however, a consistent data set is required to make optimal predictions across a program. While comparisons in terms of program efficiency, *i.e.* the time or resources to an optimised lead or candidate nomination, are generally not available the same concentration ranges are required to create consistent, high quality, *ML-grade* data sets. The choice of the number repeats will also influence the accuracy of the measurement and hence influence predictions, which further needs to be taken into account. Overall, the number of concentration points and number of repeats will highly influence the data quality and should be a key consideration of the program if ML is to be employed and good predictions are sought.

|  |  | Number of variants | Technical repeats | Readout | z' |
|---|---|---|---|---|---|
| Primary (1-2 point) screen | Function in human assay | 1,000-30,000 | 1 | Rank order | >0.4 |
| Confirmation (3-5 point) screen | Function in species relevant assays | 100's | 1-3 | Approx $IC_{50}$, ±SE | 0.5-0.7 |
| Profiling in vitro (12 point) assay | Affinity | 10's | 3 | KD | >0.9 |
|  | Engogenous cell based function |  | >3 | $IC_{50}$, ±SE | >0.8 |
|  | Developability |  | 3 | Risks ranked |  |
| Profiling in vivo | Efficacy | 1-5 |  | PD endpoint |  |
| early lead to optimise or final candiate selection for FTiH |  | 1-3 |  |  |  |

**Table 1.** An illustration of the data generation from a typical assay cascade. The details of throughput, repeats, data quality assessed by Z-prime and end points will be dependent upon aspects of the platform and biology of the target.

**Box 1.** In the context of enabling ML important aspects of data generation include:

- Agreed upon *standard protocols* (including plate layouts) for all assays if data is generated internally.
- Capture data with minimal manual handling and store it in a FAIR way [33]. For early Research a lower level than FAIR data may be sufficient but is not generally advised. This encompasses agreed data access, data field descriptions, definitions and referenced controls. However, we generally recommend data beyond the FAIR data standards.
- If pre-existing data, such as public data, is used, check the origin (lab), assay type of the data set, assay variability, readout, and units of measurement and confirm that these match. Apply stringent filter criteria to remove data that is not suitable for ML. Check if there is still enough data to train a model [34–36]. A general guideline is that there should be more data points than model parameters in order to obtain reasonable performance and avoid overfitting and poor generalisation.
- Ideally establish data and processing lineage and versioning, so that source data or processing changes can easily be tracked by all teams.
- Employ at least technical repeats on the plate controls for assay QCs (Robust Z-prime & variance of repeats) and tracking of assay behaviour (*e.g.* drift). Robustness and reproducibility of the assay protocol needs to be ensured/validated beforehand.
- Deploy clear acceptance criteria for data quality before use in modelling.
- For data of multiple classes (*e.g.* two classes in case of a QC method: fail/pass), assure that data points are available for each class [37].
- For regression, check that there are sufficient real value data points. If there are too many data points with a cutoff (*e.g.* pIC50 > 100 uM), there might not be sufficient data to train a regression model and only a classifier might be possible.
- If real valued data (not multiple classes) is used, data should include a wide range (*e.g.* pEC50 values of 7 to 11).
- Clear business rules for the (non ML) pre-processing or filtering of the data, for example for curve fitting or cut-offs for filters (*e.g.* purity).
- Deployment of a set of controls constantly throughout a program to enable appropriate normalisation. Unlike for small molecules, for biologics, controls are needed to monitor both the "performance" of the assayas well as process parameters along the whole value chain. This allows for monitoring parameters like changes in quality of DNA used to produce the clones/mAbs or expression artefacts (host cell performance, cell batches, etc.) among other aspects.
- Tracking of relevant confounding variables through metadata.

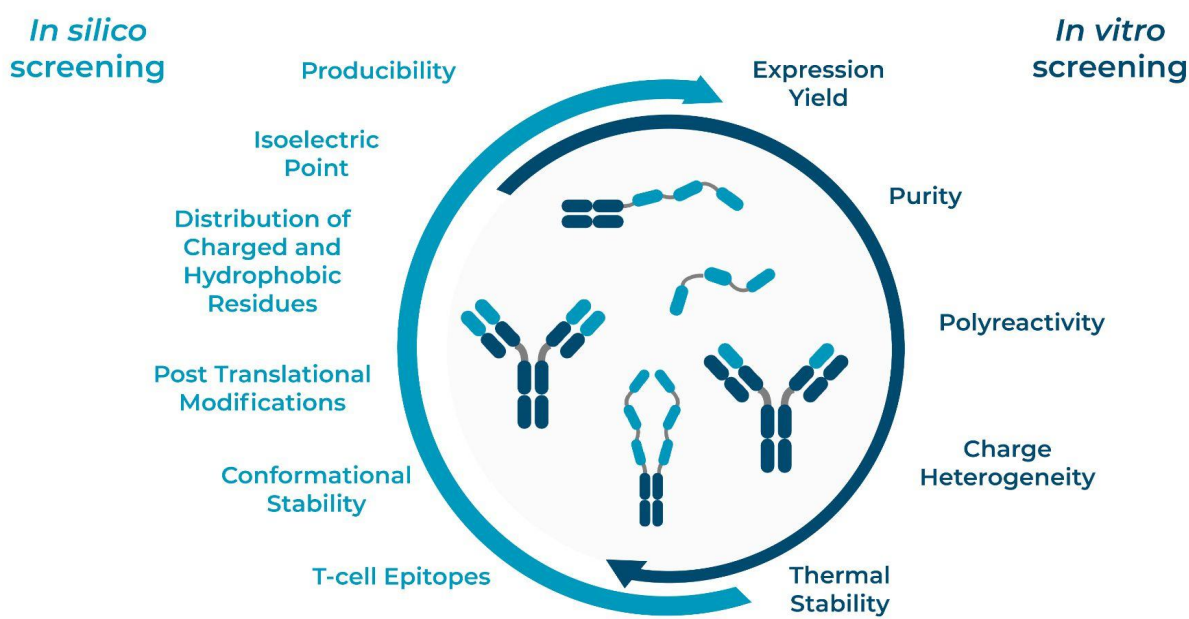

**Figure 4:** Key properties that are typically co-optimised to achieve a specified developability attributes of the CDTP.

Assay considerations are equally important for developability attributes. Developability encompasses the ability to manufacture, formulate and package a stable, homogeneous, high concentration, specific drug as cost and time efficiently as possible [15,38]. Being able to predict the developability profile of a research molecule from sequence and HT assay data would ensure molecules with poor developability are either removed from the R&D workflow or the detrimental property is re-engineered. When considering both pharmacology and developability attributes for molecules that are not simple IgGs, such as bispecific or multivalent formats, the individual building blocks may behave differently in the final format. Therefore, screening in the CDTP-envisioned product format as early as possible in the assay cascade is recommended [39]. A range of commonly optimised or screened properties are captured in Fig. 4.

## 3. Components of a (good) ML process

All applied ML requires process validation. Process validation, unlike model validation, is crucial since we need to validate that the entire process - from data collection and processing to the actual model predictions - is applicable to the given problem. We define the process of deploying ML models in a biotherapeutic environment to create actionable insights as everything that is required from the data creation to the model building and making predictions. The main steps are laid out below.

Process validation plays a crucial role in any ML application for two primary reasons. Firstly, it ensures a realistic assessment of model performance, which is vital for determining the effectiveness and applicability of a model within a real-world drug discovery setting. Secondly, process validation enables the reliable comparison of different ML models and data processing steps, allowing researchers to identify the most suitable techniques for specific tasks.

*A good ML process consists of the following steps:*
1. **Data collection:** Obtain diverse, high-quality, and relevant data from relevant sources including literature, public and commercial databases, and proprietary wet-lab experiments. Particular care needs to be taken when mixing data sources due to high variability in execution and data collection standards, in particular when dealing with human-labelled training data [40].
2. **Data curation, pre-processing and standardisation:** Clean, organise, and transform the data to ensure consistency and reduce noise. It is helpful to adopt similar practices to the ones that have been established for QSAR models [41,42], that is, assure the data is collected using standardised experimental protocols and/or molecules with common formats.
3. **Exploratory data analysis:** Examine the data to understand its characteristics, imbalances, and distributions before building complex models. Collaboration with experimental scientists is essential here for deeper understanding of the data [43]. In particular, data scientists need to acquire domain knowledge and understand the science behind the data. For example, while connecting 'microscopic' molecular properties of therapeutic antibody candidates with the 'macroscopic' biophysical measurements, it is crucial to develop an understanding of computational biophysics, protein structure and dynamics as well as the experimental biophysics protocols.
4. **Choosing a model performance metric:** Select appropriate metrics that align with the nature of the data, and the application [18,44–46]. Note that metrics might be different in specific applications such as protein structure prediction [47,48] from say prediction of biophysical measures associated with expression, purification, conformational stability and colloidal interactions of proteins.

5. **Model components and model choice:** Determine the most suitable model based on the complexity of the problem, the available data, and other constraints [18]. In certain cases, mimicking biological processes such as generation of unwanted immune responses against the therapeutic antibodies, it may be essential to develop multiple models each representing a specific step in the process and tie them together to obtain an improved understanding of the process itself.
6. **Evaluation:** Assess model performance using techniques like cross-validation, different data splits, and comparison to baseline methods to understand realistic model performance. This involves the training of different ML methods and benchmarking their performance using the same evaluation criteria. In some cases, consensus predictions or other ways of ensembling the models can perform better than individual models developed using specific ML methods.
7. **Putting the model into production:** Ensure scalability, computational efficiency, compatibility, interpretability, and monitor the data and model performance over time in a production environment. This might require the repeated validation and deployment of models as new data arrives and hence the building of strong ML ops pipelines.

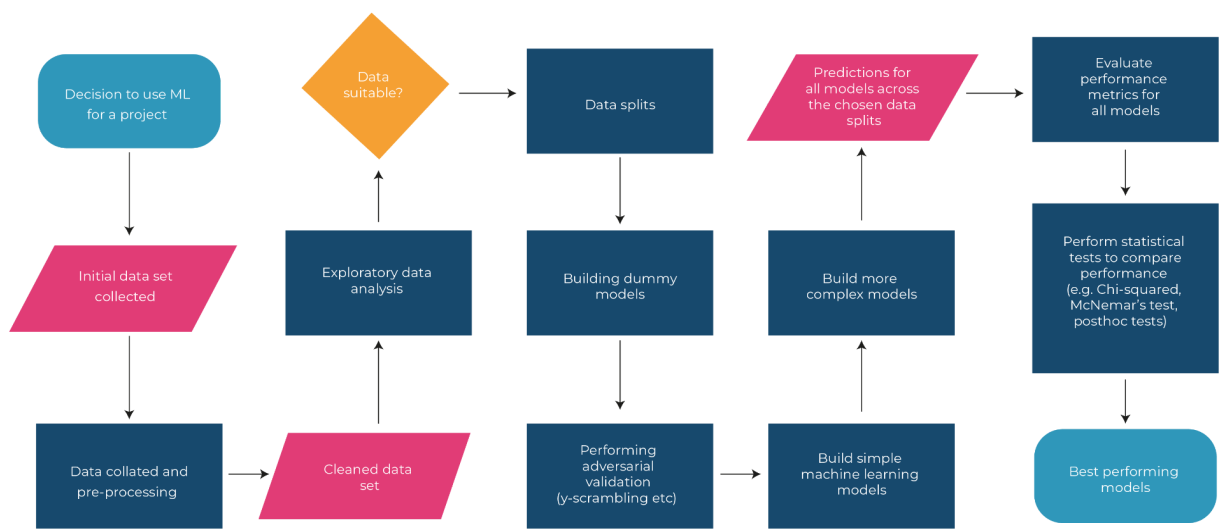

**Figure 5**: A systematic process overview of the different steps that are required to evaluate ML models in antibody discovery. These steps are generic and apply to most ML processes.

In order to ensure realistic performance assessment and the biggest impact to drug discovery projects, ML researchers and engineers need to adhere to best practices for all of the above mentioned steps. Doing so will result in more transparency and more reliable performance in real-world ML applications. The full process is summarised in Fig. 5.

In the following we will discuss each step of this process outline, highlight potential pitfalls, and make recommendations for best practices.

**4. Data collection**

In this section we delve into the essential aspects and considerations of gathering high-quality data for ML in a drug discovery setting. The goal of this step is to ensure that the collected data is accurate, relevant, and suitable for training ML models to make reliable predictions. This section covers important topics such as predictive validity of assays, data correctness, choice of measurement metrics, handling biological variability, data normalisation, addressing challenges in drug discovery data, and detecting and dealing with data drift. We will address each of these topics separately.

   a. *Predictive Validity of Assays*: Predictive validity of an assay refers to its ability to predict a desired outcome accurately [25]. In this case, we refer to it as the likelihood of an assay outcome translating into a more complex experiment, which could be a more complicated assay or even a clinical trial. It is crucial to validate that the endpoint that is used correlates with the desired outcome whenever proxy assays are used (*e.g.* endpoint/prediction needs to be correlated with go/no-go criteria). For example, a biochemical antagonism assay should first be validated to translate into a more complex cell-based read out. This is essential to guarantee that the optimisation process guarantees useful candidates. While this is mainly the responsibility of the experimental domain expert, it is crucial that the entire team is aware of the limitations that are imposed on ML models by the data. If the data is not predictive, then the model will not have any impact on the program. For example, if biochemical assay data is used to predict binding, then one cannot expect the ML model to reliably predict activity in a cell-based assay if the correlation of one with the other wasn't validated upfront.

   b. *Determine the Correct Set-up of the Assay*: As previously discussed, deciding the number of repeats and type of assay that is used will have a large impact on the data quality and quantity. Depending on the stage of the program the correct setup should be chosen and maintained throughout the whole program. We further recommend establishing, maintaining, and adhere to versioned business rules for all data (pre)processing. These should encompass normalisation, transformation, and scaling processes for the data. While the majority of this work will be the responsibility of senior experimental colleagues, it is crucial here also to get input from the data science colleagues to understand, for example, which consistent set of controls enables the best pre-processing of the data for ML purposes.

c. *Confirming Data Accuracy*: Ensuring data accuracy involves proper assay calibration and reproducibility to confirm that measurements reflect the desired behaviour. Maintaining consistent conditions is critical in ML, as later changes in the assay can introduce inconsistencies and confuse the model, resulting in poor performance. Use of control molecules and regular quality assessment of the assays is advisable. Furthermore, the collection of metadata can help also improve the quality and understanding of the collected datasets [49,50]. In particular, regular visual inspection through *e.g.* box- or scatterplots of the data allows to spot errors which can then be subsequently incorporated as automated quality controls. For example, if a small percentage of values are offset from the main distribution by a consistent amount (*e.g.* 3 log units), it suggests those values have a systematic error like incorrect units conversion and a secondary peak or wrongly used threshold placeholders (like "<10 μM" which was misinterpreted as exact values).

d. *Choosing the Correct Measurement Metric and Process*: Selecting a measurement metric that is appropriate for the specific ML problem (*e.g.* AUC, EC50, maximum activation/inhibition values) and process (incl. data) directly impacts ML performance and the suitability of combining data from different sources [40]. Consistency across data processing steps and protocol standardisation is key for optimal model training. Input from data science colleagues must be sought to inform these decisions but ultimately should be driven by experimental experts. For example, when AUC values from different sources are used it is essential to confirm that baseline subtraction, hook-effect removal, or curve fitting processes have consistently been applied to the data in order to ensure comparability.

e. *Minimise Biological Variability*: Biological data often exhibits inherent variability and noise, making it important to rely on biological and technical repeats. Understanding variance between repeated measurements helps set a baseline for the best-case model performance, as a model cannot realistically outperform assay accuracy [51,52]. This can be done by using the experimentally measured error distributions to simulate repeated measurements for a larger amount of data and then evaluate different correlation metrics based on the resulting simulations as done by Brown *et al.* [53]. The impact of experimental errors is usually more significant for datasets with a limited dynamic range and is less problematic for larger ranges. For example, when the error is nearly half of the dataset's dynamic range, then achieving a meaningful correlation becomes nearly impossible. This implies that for early projects (*e.g.* hit finding) it poses usually a small problem as for example in late stage lead optimisation or when we want to step-by-step increase the selectivity in each campaign by a small amount. Notably, the variability will vary between different assays. For example, variabilities in cell-based assays tend to be around 20-30% while biophysical measurements such as melting temperature can be accurate to 0.1 degree when done with Calorimetry and hence usually exhibit much lower variability. Additionally, controls can be used to discard readouts that exhibit excessive variability.

f. *Use Data Normalisation Based on Controls*: Data normalisation using controls is essential when data comes from the same laboratory, as it helps to calibrate data, particularly for cellular assays. For example, normalising AUC measurements using a combination of plate controls and average control AUC ensures consistency (Fig. 6) or normalisation of expression data across different plates, batches, and days (Fig. 7). However, this can be challenging when using public data sources or data from different service providers, which may lack the appropriate controls, harmonisation of experimental conditions, and protocol standardisation are again key to build larger consistent data sets and using controls to normalise data sets to a common reference is essential to improve consistency of the data [54].

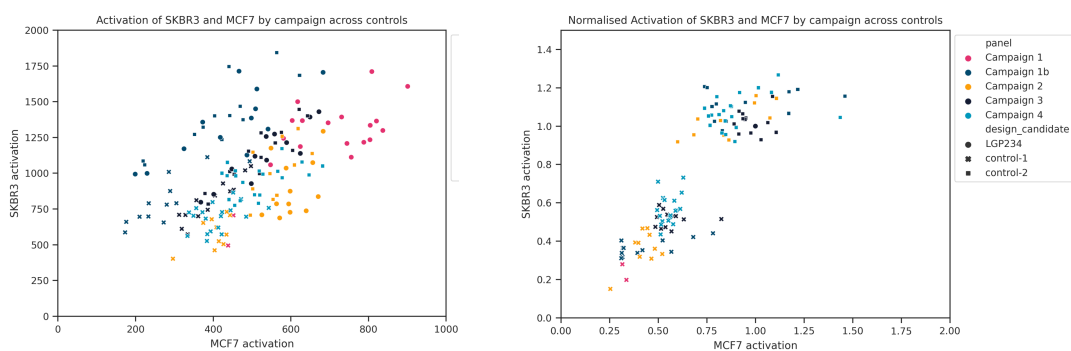

**Figure 6**. Data from different controls evaluated in cell-based assays from different campaign pre (left) and post (right) normalisation can dramatically change the picture. Here we clearly see two clusters (control-1 and control-2) emerging after normalisation.

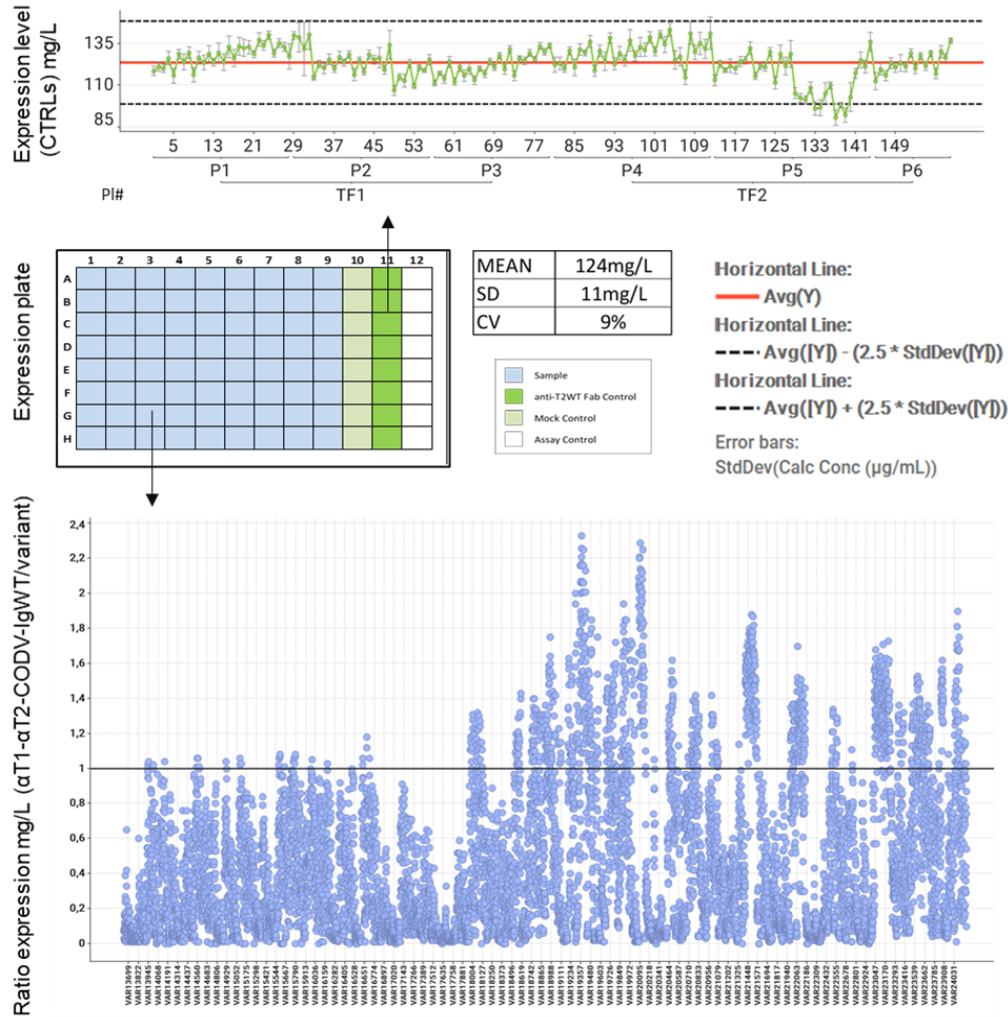

**Figure 7.** Illustration of transfection process of 12k multispecific antibodies in 96–well format (149 plates). Variants were transfected over two days (TF1 & TF2) using three cell batches per transfection day (P1-P6) level. Each plate holds 8 replicates of a transfection control. Upper panels show the arithmetic mean of the expression titers of the transfection control calculated from 8 individual replicates per 96-well plate (green in plate layout, middle panel). As the expression level of the transfection controls varies over the complete transfection process (e.g. due to changes in cell batches), it is required to normalise the transfection titers of the sample molecules (blue in plate layout, middle panel) against the mean transfection levels of the transfection controls on each plate when aiming for an overall comparison of the transfection levels of sample molecules distributed over all 149 sample plates. Normalised expression ratios are shown in the lower panel. Source: [29] (supporting material)

g. *Other Challenges with Drug Discovery Data*: Drug discovery data often has set cut-offs, such as maximum concentration limits in concentration-response curves. They are derived from the specific targets of the program (*e.g.* the required target potency) and will cover a reasonable range around these targets. For example, concentration-response curves have a number of dilution steps and will hence cover a few orders of magnitude of potencies. While these values are usually set to the most relevant range for the program, the properties of the molecules might not always fall within these ranges, leading to values beyond the limits. For example, for an assay limit of 100 uM on the upper side, a compound might hence receive a cut-off value of >100 uM.

Such cut-offs can restrict the ability to train regression models, requiring the use of categorical models (classification) with lower resolution predictions (for example >5uM or <5uM potency). Therefore, it is crucial to understand the type of data to collect for maximising the model's ability to make relevant predictions, especially when relying on public data sources [55–57].

Another challenge with drug discovery data is the harmonisation of experimental data across different functional units of an organisation. For example, the data on protein expression, purification and biophysical characterization performed at pre-formulation stages may not be easily translatable to the similar experiments performed at the CMC development and formulation stages. This often happens because of several reasons including the cell lines used (transient expression versus CHO cell line optimised for the candidate), scale of protein made and differences in the experimental protocols followed in the discovery and development stages and even differences in the instruments used to make the experimental measurements. This ties in with the translatability of early stage proxies with late stage measurements that we have touched on earlier.

A third challenge in experimental data collected in a discovery setting is time lapse. Typically, it is harder to use data on older projects than the current ones, even if the same target is being worked on again due to change of data collection and storage methods. This also means that the experimental data collected and ML models built a couple of years earlier may not be current because of adoption of newer instruments and experimental methods by the experimentalist laboratories. Therefore, collaboration between data scientists and experimentalists needs to be on a continual basis to fully realise the potential of digital transformation in the pharmaceutical industry.

h. *Detecting and Dealing with Data Drift*: Being aware of changes in experimental setups, such as reagent changes, equipment accuracy variations, or data processing alterations, is vital, as these can have significant impacts on the ML models trained. Consistently monitoring and addressing data drift, *e.g.* via the deployment of a appropriate set of controls, helps maintain model performance and ensures that any changes in experimental conditions do not go unnoticed. In addition, periodic updates of the model using the greater amounts of available experimental data is also recommended to minimise the impact of data drift. We generally recommend the visualisation of process data alongside molecular profiles throughout the program.

In case of a failed quality control, it is necessary to diagnose the source of error. Here metadata can also be helpful.

**5. Data curation and pre-processing**

Antibody engineering by means of ML mandates stringent data handling and preprocessing, key to attaining reliable, meaningful outputs. Outliers, noisy data, or data that might confuse a model (for example pIC50 of 5 in classification to distinguish more clearly between active and inactive molecules), should be carefully addressed. Although reducing data quantity, removal of data can be beneficial for the predictive performance of a ML model.

Data curation and preprocessing has a major impact on the behaviour and performance of a ML model. The Cheminformatics community has over the years established clear guidelines on how to process and filter data to make it suitable for ML purposes [41,42,58–61].
While there are some attempts to provide similar guidelines for bioinformatics and ML for biology [62], most publications are limited to best practices for computational modelling, but do not offer advice for data preparation and quality [17,22,63–68]. For data preprocessing of antibody and T-cell receptor data, the AIRR (Adaptive Immune Receptor Repertoire) community has established guidelines that enable standardised input and and output of AIRR data, which enables the communication of AIRR-compliant bioinformatics tools [69–72].

In the following we highlight key steps that should be taken in order to obtain good predictive performance and to ensure that this performance is representative for the actual application in a program. Data integration, cleaning, and binning is typically performed without the need to look at the ML model, while data transformations, feature extraction, selection and transformation are performed repeatedly during the model training and evaluation process to evaluate the model performance in various settings.

    a. *Data Integration:* Many large company databases and public data sources are not ideal for training ML models due to the lack of appropriate controls required for normalisation and varying or changing assay protocols over time. Generally, better data quality results in better model performance. Nonetheless, training models on combinations of different data sources, for example public and in-house or different public data sources can be beneficial under certain circumstances [73–75]. Kramer *et al.* ([73–75]) concluded based on a large-scale analysis of small molecule (ChEMBL) data that augmenting mixed public IC50 data by public Ki data does not deteriorate the quality of the mixed IC50 data if the Ki is corrected by an

offset. However, in our experience this is not the case for biologics and consistent internal data leads typically to the best outcomes since the production process has a much bigger impact on the final molecule and variations in it hence unproportionally affect the results. In order to decide how much data you need, a common rule of thumb is at least ten times as many data instances (data points) as there are data features. However, this depends strongly on the selected features, the quality of the data, and the complexity of the problem [76]. There are hence many exceptions to this rule.

b. *Data Cleaning:* Enhancing model accuracy can be achieved by removing noisy data points or outliers. For example, when training a classification model, excluding data points with pIC50=5 (commonly considered the threshold for inactive compounds) can improve model performance. This step may involve techniques such as outlier detection, data imputation, and standardisation of units and formats. It is in general advisable to get input from experts who understand the experimental setup and data very well when performing this step. It is for example common in many databases that active compounds are over represented due to reporting biases. For example [77] reported that the average pIC50 value of the whole distribution of data in ChEMBL25 is 6.57 for small molecules. This is likely different to prospective unbiased library screens, where it is common for 95% of the compounds to have pIC50<5, a large train-test distribution shift. For biologics these ratios will vary but being aware of such changes in the distribution is equally important in order to allow adequate processing (*e.g.* resampling) the data to create a more representative data distribution for the actual application.

c. *Binning of Data*: For classification, it may be necessary to bin data into active and inactive groups. The choice of group ranges can significantly impact model performance. Initial surveys of the data as well as feedback from the experimentalists can help set appropriate data bins. For example, IgG antibody purification data is often obtained using size exclusion chromatography (SEC). The experimental result consists of a chromatogram showing relative abundance (peaks) for extents of high molecular mass species, monomer content and low molecular mass species in terms of the percentages of the areas under the peaks. Percent monomer is often used as an indicator of quality of the antibodies when these measurements are performed over a large set of them. The range of percent monomers in these samples may vary from 0 to 100%. A typical quartile-based binning of the data without any inputs from the experts in the field may bin the samples as of poor quality (Percent monomer below 25%), good quality (25%-75%), and high quality (>75%). However, domain experts would often consider IgG antibodies with percent monomer below 90% as of poor quality, 90-95% as of good quality and those with >95% monomer content as of high quality [78,79].

d.  *Averaging Data Over Technical and/or Biological Repeats*: Data is typically averaged over technical and biological repeats, helping reduce noise and error. If the data distribution is narrow, averaging is advisable. However, if data points are disparate, outliers are better discarded, unless a clear rationale for their preservation is presented. The trade-off between diversity, replicates, and the final number of compounds tested requires careful consideration as previously discussed. Based on our experience, averaging repeated measurements and using the mean to train ML models is a best practice. However, individual repeats might be more suitable for certain models, especially when data errors or noise levels are used for calibrating model uncertainties as is commonly the case for Bayesian methods [32,80].

e.  *Feature Extraction and Selection*: Converting raw data into a set of features or descriptors is crucial for effectively training ML models in drug discovery. In the antibody or protein engineering space, features may include protein sequence or sequence-derived structural features, designs (*e.g.* a combination of combinatorial antibody parts), 3D structure, learned representations, or more conventional molecular descriptors (amino acid composition, dipeptide composition, tripeptide composition, or pseudo amino acid composition). Physicochemical properties of amino acids, such as hydrophobicity, charge, size, and polarity, can also be used to compute various descriptors, including autocorrelation, Moran, and Geary coefficients. Finally, the incorporation of evolutionary information, for example through multiple sequence alignment, has been particularly helpful for computational structure prediction.

    For protein sequence-based features, values can be directly extracted from the sequences, such as amino acid type, evolutionary information (*e.g.* profile representations in the form of PSSMs from PSI-BLAST), or features predicted by other tools, such as secondary structure and solvent accessibility. Representation models like transformers can be used to learn features from large volumes of unlabeled data, aiming to represent the innate structure of the data. Several pretrained representation models are available for proteins (*e.g.* CPCProt, deepGOCNN, ESM-1b, ProtTrans/ProtBert, rawMSA, SeqVec, GearNet, or UniRep, AntiBerty and AntiBerta) and we recommend carefully evaluating these depending on the specific task at hand [34,81–86]. In addition to these, physicochemical features for individual amino acids represented in AAIndex may also be useful [87–89].

    Feature selection is important when dealing with protein sequence-based features as many features can be redundant. Feature selection offers several advantages, including a decrease in the overall number of tunable

parameters in the algorithm, reducing the likelihood of overfitting. A reduced number of input features can also increase the algorithm's speed, which is crucial for large-scale applications. Most importantly, a concise list of relevant features aids in understanding the essential characteristics of the problem at hand. Feature selection can be categorised into three types: (i) 'wrappers' use the ML algorithm as a black box to select features based on their performance, (ii) 'filters' select feature subsets without considering the ML algorithm, and (iii) 'embedded' techniques are part of the ML algorithm training procedure. Choosing the correct representation for the task at hand is crucial and typically more important than the choice of the model used. It is recommended to evaluate a range of representations in combination with simpler models to identify the most suitable approach for the specific problem. In drug discovery, simple models have repeatedly been shown to outperform more complex ones [90–92], and should hence be used at least as a baseline before moving on to more complex ones such as deep learning. Non-redundancy can be a potential way for feature selection. This can be done by clustering the features based pairwise correlation among them for a given set of proteins/antibodies. In recent analyses of marketed and clinical stage biotherapeutics, it was shown that only a few features were needed to build profiles of antibodies likely to reach or succeed in the clinic [93,94]. Depending on the prediction target there are also many choices for the output representation. For example, one might want to predict the structure of a protein via all atom positions or a reduced representation such as the ɑ-Carbon only. Choosing the right output representation is equally important for a successful application.

f. *Feature Scaling*: Real-world datasets often contain features that are varying in degrees of magnitude, range, and/or units. In order for ML models to interpret these features on the same scale, we hence need to perform a step called feature scaling that involves standardising the range of feature values to ensure that no single feature dominates the model. Common techniques include normalisation (scaling features to a range of 0 to 1 or -1 to 1) and standardisation (scaling features to have zero mean and unit variance). Feature scaling will likely depend both on the data and the algorithm that is used. Some algorithms require or preferably operate on scaled features [95,96].

g. *Data Transformation*: Apply transformations to the data to reduce dimensionality, enhance interpretability, or improve model performance. Examples include principal component analysis (PCA), t-distributed stochastic neighbour embedding (t-SNE), and log transformation. Dimensionality reduction methods can be used to reduce the number of features and hence reduce training times. They can also be used to assess the importance of features, sense checks, or to confirm biological hypotheses or act as regularisation [97,98].

h. *Simulations*: there is a lack of large-scale ground truth experimental data. This hinders the development and benchmarking of robust and interpretable ML approaches [99,100]. To address this problem, there is a need to complement analyses on experimental data with simulated ground-truth data. The challenge is to generate simulated data, such that it incorporates key features observed in experimental repertoires that render ML problems challenging. Simulation frameworks for antibodies range from VDJ-recombination-like antibody generation [101,102] to synthetic antibody-antigen structures [103]. Together, these tools allow for large-scale, high-throughput and real-world relevant synthetic data generation. The here cited simulation tools have been tested for nativeness vis-a-vis experimental data. Extension of experimental observations via simulations can also help explore deeper correlations among different attributes such as aggregation and immunogenicity of antibody-based therapeutics. For example molecular dynamics trajectories could be used as input to machine learning models to enable better prediction of molecular properties such as binding, similar to small molecules [104,105]. Availability of curated publicly available datasets is required for comparison of methods based on a single agreed-upon datasets [13]. The OAS (Observed Antibody Space) [106] and iReceptor databases [107] represent starting points where novel datasets that are associated with function metadata may be integrated. These datasets could be set for different ML tasks ranging from antibody structure prediction, antibody-antigen docking to antibody developability prediction. These datasets would not only represent a *data standard* but also necessary building blocks for public competitions [48,108–111] . Competitions are integral to mapping both those areas where predictability is good as well as where knowledge blank spots exist.

## 6. Exploratory data analysis

Exploratory Data Analysis (EDA) is an essential step in the ML process. It allows for a better understanding of the data and helps inform subsequent modelling decisions. In this section, we discuss several aspects of EDA in the context of drug discovery:

a. *Assessing property distributions*: Analysing the distribution of biophysical properties, activation (*e.g.* IC50 or Ki value range), potency, selectivity, positional amino acid frequency, antibody topology, or other relevant features in the dataset. This analysis can help identify potential biases, outliers, or trends that may impact model performance. It will also impact the model's prediction ability and enable determination if additional preprocessing or normalisation steps are required.

b. *Coverage of the target space and dynamic range of the data*: Coverage of the target space (the final outputs of inputs) and dynamic range of the data: Evaluate the data's coverage of the target space to ensure that the model can make accurate predictions across the entire range. Assess the dynamic range of the data, as small ranges may lead to poor model performance. For example, if selectivity ranges are small it is unlikely that the model can make reliable predictions far outside these ranges (c.f., model applicability domain). When dealing with very small dynamic ranges, the question of sufficient resolution of the corresponding assay might arise. If the underlying assay is not able to distinguish variants with diverse properties in a significant manner, computational methods built on top of such data will most likely fail as well. The dynamic range of a dataset can have a large impact on the apparent correlation between experimental and predicted activity and the literature is full of examples of what appear to be impressive correlations on datasets that span an unrealistically high range. So when testing a model it is also important to ensure that the evaluation range is representative of the application it is intended for. When data within this typical range is considered, these apparent correlations can decrease dramatically [112].

c. *Evaluating data imbalance*: Assessing the balance between different classes or ranges of values in the dataset, as imbalanced data may negatively impact the performance of ML models. Techniques such as re-/oversampling [113–115], undersampling [113–115], changing the decision threshold (for classification) [116], or using weighted loss functions [115] can help address this issue and should be considered where appropriate [117–120]. Furthermore, different metrics should be chosen for the performance evaluation depending on the data imbalance. We will discuss this further in the later sections on metrics.

d. *Model applicability domain*: It might be possible to evaluate the applicability domain of the model based on similarity of training to test/production data or uncertainty [121,122]. For similarity, consider factors such as the similarity of the training set property range to the target property range, input sequence similarity, or clustering representations. Keep in mind that sequence similarity does not always imply phenotypic similarity, as sequence-similar antibodies may bind different antigens. The applicability domain should in particular be considered when deploying generative models [123], since these can easily exploit weaknesses of the scoring functions [124,125].

e. *Simple correlation and cluster analysis*: Perform correlation and cluster analysis to gain insights into the relationships between variables, identify potential outliers or patterns, and inform the choice of features and models. This information can be valuable for improving model performance and interpretability. Additional methods for outlier detection should also be considered, *e.g.* [126].

**7. Choosing the correct model loss function and performance metrics**

Selecting a set of appropriate metrics is critical for assessing the performance of ML models in drug discovery [127]. This section will discuss various metrics for regression and classification tasks and when to apply them. Rather than being comprehensive we only try to give the reader a flavour of the variety of methods and refer to literature for further reading. It is important to note that metrics are different from loss functions. Loss functions measure the model performance during training and are used to optimise machine learning models by minimising the loss in order to derive the optimal performing model. The loss function hence usually needs to be differentiable (*i.e.* a gradient can be calculated) with respect to the model's parameters. Metrics, on the other hand, are used to monitor and measure the performance of a model both during training and testing. Not all metrics are differentiable, in particular if they are not used for training but only for the final evaluation of the model. While there is a substantial overlap between the two, here we focus on the model performance metric [128]. For the final model evaluation we typically look at multiple different metrics, while for the loss function typically a single one is chosen where additional terms may be added for regularisation [129]. Both need to be chosen carefully alongside the optimiser in order to be successful. The advice around the metrics that are discussed below will generally hold up for the loss function as well as for the final model evaluation.

1. Regression Metrics: It is crucial to evaluate regression models using multiple metrics, including Pearson and Spearman correlation coefficients and Kendall's tau. Combinations of different metrics can provide insight into model performance as they highlight strengths and weaknesses in different regimes. Examples for common metrics are [18,20,130]:

    - Mean Absolute Error (MAE/L1): This is the absolute value of the difference between the prediction and the observed value. It should be used when you want to minimise the effect of outliers, as it is less sensitive to these than Mean Square Error (MSE).
    - Mean Square Error (MSE): This is the square of the absolute value of the difference between the prediction and the observed value. The MSE is on the other hand susceptible to outliers (extreme values), and these might impact the model performance unproportionally. MSE may also punish errors too heavily if your targets have a large value spread. Mean squared logarithmic error may then be more appropriate, specifically if you're dealing with large, unscaled quantities.

- R-Squared: Use when you want a robust and interpretable metric that considers the proportion of variance explained by the model.
- Pearson Correlation Coefficient (r): Use when assessing linear correlations between two sets of data.
- Spearman Correlation Coefficient (rho): Use when assessing non-linear, monotonic relationships between two sets of data, specifically to measure rank correlation.

It's essential to remember that correlation coefficients have themselves an associated error, dependent on both the correlation coefficient itself and the number of data points that are used to obtain it. It is hence good practice to evaluate confidence intervals for the correlation coefficients, in particular when comparing these across different models. Only if two correlation's confidence intervals do not overlap, there is a clear advantage. However, existence of an overlap does not necessarily imply a lack of difference between two models [131].

2. Classification Metrics: For classification tasks, it is important to select the appropriate threshold for optimal performance. Several metrics can be used to evaluate classification models:

   - Balanced Accuracy: Use when dealing with imbalanced data, as it considers both sensitivity and specificity.
   - Accuracy: Use when the problem is balanced, and classes are equally important.
   - Precision: Use when you want to focus on the accuracy of true positive predictions.
   - Recall: Use when you want to focus on the proportion of true positive predictions out of all possible positive predictions.
   - F1 Score: Use when dealing with imbalanced data and focusing on the positive class. It balances precision and recall.
   - ROC-AUC (Receiver Operating Characteristic/Area Under the Curve), also called AUROC: Use for balanced data, as it measures the trade-off between true positive rate and false positive rate [130]. Avoid using ROC-AUC for imbalanced data [132]. ROC-AUC can summarise the performance, with perfect classifiers having an AUC of 1 and a random one having 0.5. It's another measure of model calibration as it assesses model performance across all possible decision boundaries and is directly related to the Mann-Whitney statistic [133].
   - Partial AUC: Instead of the area under the entire curve this is a restriction to certain sensitivity and specificity ranges. This might be useful when certain decision boundaries are irrelevant.
   - PR-AUC (Precision-Recall Area Under the Curve): Use for imbalanced data sets or when you care more about the positive class, as it focuses on precision and recall.

- MCC (Matthews Correlation Coefficient): Use when you want a metric that considers all four values of the confusion matrix, providing a comprehensive evaluation of binary classification.
- Cohen's kappa: This is the chance-corrected accuracy. This accounts for class imbalance and is useful for determining if a model is better than a naive Bayes model.

For multiclass predictions, metrics such as Matthew's correlation or extensions of the above scores can be used but require additional care [134,135]. Choosing the metric with care is crucial since these can otherwise be misleading. For example, Cohen's kappa should be avoided [136]. In addition, when predicting probabilities (rather than classes) other metrics such as the Kullback–Leibler divergence need to be chosen [137].

When selecting a metric, consider the specific requirements and goals of the drug discovery project, as well as the distribution of the data. Different metrics may be more suitable under different circumstances, such as data imbalance [138] or the importance of certain classes. For example, high specificity or low false positive rates are often desired, especially in scenarios where the rarer positive class is the main interest. In general it is advised to always assess models for multiple metrics in order to find the most suitable one for a given task. In most common packages, such as sklearn [138,139], all metrics are already pre-implemented and hence readily available. For classification, it is also helpful to change the threshold for the classification based on the task at hand as the standard binary classification threshold of 0.5 is not always ideal.

It is good practice to monitor the learning curve over the training process for both the loss function and the performance metric. This allows for spotting undesired behaviour such as overfitting.

## 8. Model components and model choice

Choosing the right ML model for drug discovery projects is crucial for achieving desired results. This section discusses various aspects of ML models, from data considerations to model types and best practices.

1. General Data and Program-Specific Data: Depending on the dataset size and problem scope, you can use pre-trained or multi-class models or train new models from scratch. When you have a large, diverse dataset that covers various aspects of the problem, pre-trained models can be fine-tuned for your specific application. For smaller, domain-specific datasets or unique problems, training a new model might be necessary.

2. Model Requirements: Consider the specific goals of the drug discovery project and the characteristics of the data when selecting a model. This includes interpretability, computational resources for training, and the model's ability to generalise to new data.

3. Data Preprocessing: Data preprocessing techniques like StandardScaler, MinMaxScaler, and PowerTransformer can be crucial for achieving good results [140–142]. StandardScaler removes the mean and scales features to unit variance, while MinMaxScaler scales features between a specified range. PowerTransformer applies a power transformation to make data more Gaussian-like.
Standardising data is important for many ML estimators because they may perform poorly if the individual features do not resemble standard normally distributed data. This can be particularly relevant for models that rely on the assumption of normally distributed data or are sensitive to feature scales, like linear regression or support vector machines.

4. Model Types: Depending on the problem you might want to use a different set of descriptors (*e.g.* structure, sequence, physchem descriptors). Structure-based models use 3D molecular structures, sequence-based models focus on the amino acid sequences, and descriptor-based models use calculated features to represent molecules. You can use different descriptors with different models and in many cases different descriptor-model combinations will result in different performance. Automating the process so that you can test a variety of combinations is hence generally useful and a best practice.
In terms of ML models, there is a range of models that could, or indeed should be tested, and in many circumstances it is advisable to start from simple and go to more complex. Below we suggest a general flow of models that can be used to first test the data splits and evaluate baselines for the model performance:
    - Dummy Models: Start with simple baselines to evaluate the performance of more complex models. The simplest models are dummy models, which for example can just make random predictions or majority class predictions. These are helpful to establish an absolute minimum baseline to beat.
    - Adversarial validation [143,144] is a technique used to assess the degree of similarity between training and testing datasets in terms of feature distribution. One way to perform adversarial validation is to train a binary classification model to predict whether a given sample belongs to the training or test dataset. The training and test datasets are combined, and labels are assigned to each sample (0 for training, 1 for test). The performance of this model, typically evaluated using an ROC curve or AUC score, gives an indication of how similar the two datasets are. If the adversarial model performs poorly, this suggests the training and test data are similar, and conventional validation techniques should work well. Conversely, if the adversarial

model performs well, this indicates that the training and testing data distributions differ significantly, potentially signalling a risk of overfitting and misleading model validation results. The goal of adversarial validation is to ensure that the model built using the training dataset will generalise well to the unseen test dataset, reducing the risk of overfitting and improving the robustness of model predictions [145]. It can be used to invalidate wrong hypotheses. A related method is y-scrambling [146]. In y-scrambling the model is first trained on the original data and the performance metric is observed. The y-labels are then shuffled so that the correct feature-target pairs are now replaced with the incorrect feature-target pairs, in other words with incorrect labels. Now the model is retrained on this data and the performance metric observed. The last step is repeated a few times to obtain a sample of the performance. It is expected that the model performs well over the original data and poorly on the shuffled data. If that is not the case and the metric doesn't vary much, then that means the predictions aren't robust and the model predictions are likely not reliable. As an example, in [147] the authors showed that the original model had a high $r^2$ and low RMSE scores which were, however, closely replicated by y-scrambled models. This immediately casts doubts on the original model's validity.
- Conventional ML Models (Random Forest, SVM, etc.): Use these models as a starting point before exploring more complex (deep learning) models. These will usually be easy and quick to implement and give you a sense if there is a signal in the data. They will also allow you to meaningfully assess any improvements of more complex models and are typically more robust.
- Deep Learning Models: Consider using deep learning models when you have a large, complex dataset, and the problem requires advanced feature extraction. Certain types of architectures, such as equivariant neural networks, can be used even if little data is available.
- Physics-based Models and Simulations: These models can be used when you have structural information and require a more detailed understanding of the interaction mechanisms. A wide range of tools like Rosetta [148,149] , ZDOCK [150], HDOCK [151] are available and should be chosen depending on the problem [152,153].

5. Best Practices:
    - Combining Models into Ensembles: Ensemble methods can improve model performance by combining the strengths of multiple models. This is particularly the case if you combine different types of models (not the same type with different initialisation), such as physics-based with conventional ones.

- Common Pitfalls and Best Practices: Always validate your models using appropriate metrics, avoid overfitting, and tune hyperparameters carefully. It is also essential to use a correct data split and avoid data leakage [154]. This should be carefully tested beforehand. Data leakage in particular has become a large problem in computational protein structure prediction [155] and needs to be carefully considered, since sequence alone is not indicative for good data splits and data leakage and protein homology also needs to be taken into account (*c.f.* section 9).

In summary, choosing the right ML model for drug discovery projects involves considering various factors such as data size, model requirements, preprocessing, and best practices. Always start with simple models and move towards more complex models as necessary, keeping in mind the specific goals and characteristics of your drug discovery project.

**9. Evaluation**

Model evaluation is a critical aspect of drug discovery projects that ensures the chosen models are effective and accurate [127,145]. This section delves into various components of model evaluation, from validation to data splits and feature analysis.

1. Process Validation vs. Model Validation: In drug discovery, every model validation is a process validation rather than just a model validation [31]. This is because the entire process, including data preprocessing, feature selection, and model training, contributes to the overall performance of the model. Therefore, it is essential to validate and optimise the entire process end-to-end to ensure the best possible results.
2. Benchmarks: Benchmarking against dummy, simple, and state-of-the-art approaches is crucial for evaluating the effectiveness of complex models. By comparing the performance of the proposed model with existing methods, it is possible to demonstrate the superiority of the new model and justify its use in the drug discovery project.
3. Metrics: As discussed earlier, choosing the correct metrics for model evaluation is critical. Ensure that the selected metrics are appropriate for the specific drug discovery problem, the model, the task, and that they also adequately capture the desired aspects of model performance.
4. Significance testing: An important part of the model evaluation is the comparison between different tested models and whether there is a significant difference in their performance. Since model performance is based on a statistical sample (the data split), evaluation of the performance differences should be performed using statistical methods.
Generally, a statistical hypothesis test quantifies how likely it is to observe the outcome given the assumption that the model outputs have the same distribution. If the result of the test suggests that there is insufficient evidence to reject the null hypothesis that there is no difference in the model, then any observed difference in model skill is likely due to statistical chance. For comparison of two models, it is recommended to use McNemar's test in cases where there is a limited amount of data and each algorithm can only be evaluated once [156] or a resampling method with 10x10-fold cross-validation with a corrected paired Student-t test [157,158]. Comparing multiple models at once requires different tests such as the Holm–Bonferroni method, a Wilcoxon signed rank test with adjustment for multiple testing or the Friedman two-way analysis of variance by ranks test (in short Friendman test) [159–161]. While it is generally recommended to use adjustments of the p-values for comparison of multiple models, there is no consensus as to which method should be used. We refer the reader to [162] for a practical tutorial, and note that looking at the overlap of error bars is insufficient [163,164].

5. Model Performance vs. program impact: While academic and research groups are mainly interested in improving the model quality, in real-world drug discovery scenarios the real impact comes from the improved quality of decision making leading to the generation of better drug candidates. The performance of a model should hence be evaluated in the context of its impact on the drug discovery program. In many cases, a few percent uplift in the model performance is irrelevant given that the noise in the data is usually high and models simply overfit. Getting a small amount of extra, high-quality data, would usually result in much better outcomes and should hence be considered as an alternative to working for weeks on new or better models.

6. Drug Discovery-Specific Metrics: There are additional metrics for ML-driven discovery processes. While these are likely less familiar for a ML expert, they may be used to evaluate the model's ability to discover new drugs. These include, for example:
   - Enrichment: The fraction of active compounds (*e.g.* binders) in a selected subset of compounds compared to the fraction of active compounds in a randomly chosen subset [165–167]. A related metric is the normalised discounted cumulative gain (NDCG), which indicates if the top predicted results are enriched for truly high performers [168]. In contrast to the Spearman correlation, NDCG weights the ranking of the top of the list higher.
   - Scaling factor: The number of compounds that can be reliably predicted per experimentally tested compound.
   - Search efficiency: How quickly a method can discover the top-performing compound in a dataset, usually used in simulations (and retrospective studies).

7. Data Splits: Defining data splits that better assess real model performance is crucial for accurate evaluation. Predictive models are not universally applicable and generally perform better when predicting the activity of molecules similar to those in the training set and worse if the molecules are too dissimilar [169]. In our experience simple data splits rarely reflect the true prospective model performance in a drug discovery program [170]. For this reason a set of different data splits with varying degrees of difficulty typically allows one to get a better sense of the true prospective model performance. For small molecules this is for example done using random splits, splits by Tanimoto similarity, Bemis-Murcko scaffold splits, and Butina clustering-based splits on extended connectivity or Morgan fingerprints [169]. For proteins, we recommend to test several of the below dataset splitting approaches:

- Types of Data Splits: Various data splits based on sequence identity can be used for protein models [19], such as variation in sequence, mutation location, number of mutations, amino acids, physicochemical properties of the sequence, and property ranges. It is also useful to use the concept of adversarial validation [143] to establish if training and test data is easy to distinguish and the splits are done well. When mixed data sources are used (*e.g.* for training and test data we could train on public data but predict internal data), this method is very useful to assess how likely a successful outcome is.
- Approaches: Consider using 80-20 or other fixed splits, cross-validation, time splits (*e.g.* campaign-based) [171] or simulated time splits [172], or cluster-based splits for data splitting. In general a variety of different data splits with increasing difficulty should be chosen. The most difficult splits are in our experience often most representative of real world performance.
- Of particular importance when splitting the data is to check for any data leakage. This can happen if proteins with high sequence identity or homology are present in the data set and it is hence essential to check the similarity between training, test, and validation sets. A typical approach that is chosen by many researchers is to use a threshold of 25–30% for sequence identity of the training set proteins to any protein in the test set. This is enough to exclude many homologous pairs of proteins, but it is well known [47,173,174] that some homologous proteins can have virtually no sequence similarity. Such challenges should be addressed as well, by invoking additional tools such as CD-Hit [175], BlastClust [176], or TESE [177]). Such reduction algorithms often rely on local sequence alignments, and multidomain proteins might lose unique domains during the reduction. Preprocessing the sequences into domains using tools like PFAM [178] can be an option as well.
However, for antibodies or related scaffolds sequence identity thresholds of 25-30% need some further consideration as the constant regions of antibodies are highly similar and diversity is mainly limited to the Fv and CDR regions. Custom data splits (as discussed in the next section) taking the program or project specific sequence diversity and design into account, are often the methods of choice.
Other types of data leakage are of course also possible, for example, the availability of a feature during training that is not available during testing or in the application. We refer the reader to the literature (*e.g.* [179])
- In antibody D&D, one typically engages either of the following data forms: (a) Mutational variants, (b) different „wild types" (no mutational relationship; directed against same or different targets), or (c) a mix of the above. For 'general' models such as for expression or stability, then (c) is the most common case, while (a) and (b) are very common for program-specific predictions such as function during lead identification (b) or optimization of lead candidates (a). For data representative for cases (a) and (c) generic splits should be avoided and other approaches should be sought depending on the application and objective. For example, training and test splits should consider meaningful distributions of mutational positions, amino acid type, and biophysical properties (polar vs hydrophobic, etc.). For data of the form (a) and (c), data leakage can become a problem and random splits should be avoided since it will likely result in overestimation of the performance and a lack of generalisation abilities. Here, one may for example split by sequence families or cluster the sequences by similarity and avoid overlap between training and test. In general we always advise to understand the data set (*e.g.* origin and purpose of the sequence designs) at hand and a clear definition of similarity and the prediction goals in order to create data splits that are representative of real world performance.
- A last point is related to the hyper-parameters. No parameter should be selected based on the test data, and this includes hyper-parameters. Examples of hyper-parameters include the number of clusters in K-means, the number of trees in random forest, or the regularisation parameters for SVMs, or number layers or neurons in a neural network. A simple approach to guard against overestimation due to choice of the hyper-parameters based on the test set is to introduce a third validation set and select hyper-parameters based on it before testing it on the test set (ideally using cross validation).
- Impact of training data composition: there is emerging evidence that the choice of negative data can impact prediction accuracy and generalisation in both antibodies [103,180,181] and TCR specificity prediction [182–185]. Therefore, care must be taken as to how negative and positive datasets are defined.
8. For each model and metric there is a tradeoff between false-positives and false-negatives. For each project and problem there will be specific choices and these needs to be informed by the overall objectives. Clearly understanding the tradeoffs will enable us to obtain optimal performance.
9. Feature Analysis and Interpretation of Results: Evaluating Feature Importance: Assessing the importance of features helps in understanding their contribution to the model's performance and refining the model. For feature significance, for example, one can use p-values to assess feature significance or other methods like GINI impurity for Random Forest models or SHAP analysis [186].

In conclusion, model evaluation is a multifaceted process that involves validation, benchmarking, metric selection, performance assessment, and feature analysis. By carefully considering each aspect, researchers can ensure that the chosen models are accurate and effective, ultimately leading to more successful drug discovery projects. Without proper testing, there is a large risk of the performance of a model not being representative of real-world performance on new or unseen data. This can in the best case undermine the trust in scientific results and in the worst case result in huge costs or harm to patients.

**10. Conclusions**

The use of ML in antibody drug discovery holds significant promise, but its real-world impact remains limited. We believe that the next few years will see an unprecedented impact of such methods on current and future drug discovery programs. This will be due to the availability of more and better data, better methods, and more robust processes. This review establishes a clear set of best practices, primarily focused on robust data generation, capture, and model building. This includes rigorous model validation, to allow for the evaluation of real-world prospective performance in therapeutic antibody R&D. By addressing pitfalls and offering clear guidelines, this review bridges the gap between theoretical advances and practical applications in therapeutic antibody engineering. The focus on practical considerations ensures that ML applications not only accelerate the R&D process but also contribute to the development of safer and more effective biotherapeutics. Overall, by adhering to best practices and robust validation approaches, the field can progress to produce higher quality antibodies, thus offering better therapeutic options and meeting unmet medical needs. Future work should continue to emphasise the importance of robust end-to-end processes from data generation and storage to model validation and deployment. We hope that the widespread adoption of standardised guidelines will drive the field forward and maximise the benefit to patients. In summary, while ML offers vast opportunities to revolutionise antibody drug discovery, its full potential can only be realised through the adoption of rigorous guidelines and best practices as we propose in this review.

**Acknowledgements**
We would like to thank Gino van Heeke and Wade Davis for detailed comments and careful review of the manuscript and Lucy Shaw for review of the manuscript and the creation of the figures for this article. We would also like to thank Marion Schneider and Eduard Kober for generating some of the figures.

**Funding**
The Leona M. and Harry B. Helmsley Charitable Trust (#2019PG-T1D011, to VG), UiO World-Leading Research Community (to VG), UiO: LifeScience Convergence Environment Immunolingo (to VG), EU Horizon 2020 iReceptorplus (#825821) (to VG), a Norwegian Cancer Society Grant (#215817, to VG), Research Council of Norway projects (#300740, (#311341, #331890 to VG), a Research Council of Norway IKTPLUSS project (#311341, to VG ) This project has received funding from the Innovative Medicines Initiative 2 Joint Undertaking under grant agreement No 101007799 (Inno4Vac, to VG).

**Conflicts of interest**
V.G. declares advisory board positions in aiNET GmbH, Enpicom B.V, Absci, Omniscope, and Diagonal Therapeutics. V.G. is a consultant for Adaptyv Biosystems, Specifica Inc, Roche/Genentech, immunai, Proteinea and LabGenius. L.W. is an employee of LabGenius Ltd., a biotech company developing next-generation antibody therapeutics. S.K. is an employee at Moderna Therapeutics, a pharmaceutical and biotechnology company based in Cambridge, Massachusetts, that focuses on RNA therapeutics. N.F. is an employee of Sanofi S.A., a multinational pharmaceutical and healthcare company.  A.B. is an employee of AstraZeneca, a patient-focused pharmaceutical company.

**References**

[1] Senior M. Fresh from the biotech pipeline: fewer approvals, but biologics gain share. Nat Biotechnol 2023;41:174–82. https://doi.org/10.1038/s41587-022-01630-6.
[2] Wang Y, Yang S. Multispecific drugs: the fourth wave of biopharmaceutical innovation. Signal Transduct Target Ther 2020;5:86. https://doi.org/10.1038/s41392-020-0201-3.
[3] Durán CO, Bonam M, Björk E, Hughes R, Ghiorghiu S, Massacesi C, et al. Implementation of digital health technology in clinical trials: the 6R framework. Nat Med 2023:1–5. https://doi.org/10.1038/s41591-023-02489-z.
[4] Paul SM, Mytelka DS, Dunwiddie CT, Persinger CC, Munos BH, Lindborg SR, et al. How to improve R&D productivity: the pharmaceutical industry's grand challenge. Nat Rev Drug Discov 2010;9:203–14. https://doi.org/10.1038/nrd3078.
[5] Schlander M, Hernandez-Villafuerte K, Cheng C-Y, Mestre-Ferrandiz J, Baumann M. How Much Does It Cost to Research and Develop a New Drug? A Systematic Review and Assessment. PharmacoEconomics 2021;39:1243–69. https://doi.org/10.1007/s40273-021-01065-y.
[6] Wouters OJ, McKee M, Luyten J. Estimated Research and Development

Investment Needed to Bring a New Medicine to Market, 2009-2018. JAMA 2020;323:844–53. https://doi.org/10.1001/jama.2020.1166.
[7] Morgan S, Grootendorst P, Lexchin J, Cunningham C, Greyson D. The cost of drug development: A systematic review. Heal Polic 2011;100:4–17. https://doi.org/10.1016/j.healthpol.2010.12.002.
[8] Kelley B. Developing therapeutic monoclonal antibodies at pandemic pace. Nat Biotechnol 2020;38:540–5. https://doi.org/10.1038/s41587-020-0512-5.
[9] Akbar R, Bashour H, Rawat P, Robert PA, Smorodina E, Cotet T-S, et al. Progress and challenges for the machine learning-based design of fit-for-purpose monoclonal antibodies. MAbs 2022;14:2008790. https://doi.org/10.1080/19420862.2021.2008790.
[10] Narayanan H, Dingfelder F, Butté A, Lorenzen N, Sokolov M, Arosio P. Machine Learning for Biologics: Opportunities for Protein Engineering, Developability, and Formulation. Trends Pharmacol Sci 2021;42:151–65. https://doi.org/10.1016/j.tips.2020.12.004.
[11] Glatt S, Helmer E, Haier B, Strimenopoulou F, Price G, Vajjah P, et al. First-in-human randomized study of bimekizumab, a humanized monoclonal antibody and selective dual inhibitor of IL-17A and IL-17F, in mild psoriasis. Br J Clin Pharmacol 2017;83:991–1001. https://doi.org/10.1111/bcp.13185.
[12] Bauer J, Rajagopal N, Gupta P, Gupta P, Nixon AE, Kumar S. How can we discover developable antibody-based biotherapeutics? Front Mol Biosci 2023;10:1221626. https://doi.org/10.3389/fmolb.2023.1221626.
[13] Mock M, Edavettal S, Langmead C, Russell A. AI can help to speed up drug discovery — but only if we give it the right data. Nature 2023;621:467–70. https://doi.org/10.1038/d41586-023-02896-9.
[14] Bender A, Cortés-Ciriano I. Artificial intelligence in drug discovery: what is realistic, what are illusions? Part 1: Ways to make an impact, and why we are not there yet. Drug Discov Today 2021;26:511–24. https://doi.org/10.1016/j.drudis.2020.12.009.
[15] Fernández-Quintero ML, Ljungars A, Waibl F, Greiff V, Andersen JT, Gjølberg TT, et al. Assessing developability early in the discovery process for novel biologics. Mabs 2023;15:2171248. https://doi.org/10.1080/19420862.2023.2171248.
[16] Bender A, Schneider N, Segler M, Walters WP, Engkvist O, Rodrigues T. Evaluation guidelines for machine learning tools in the chemical sciences. Nat Rev Chem 2022;6:428–42. https://doi.org/10.1038/s41570-022-00391-9.
[17] Lee BD, Gitter A, Greene CS, Raschka S, Maguire F, Titus AJ, et al. Ten quick tips for deep learning in biology. PLoS Comput Biol 2022;18:e1009803. https://doi.org/10.1371/journal.pcbi.1009803.
[18] Lones MA. How to avoid machine learning pitfalls: a guide for academic researchers. ArXiv 2021. https://doi.org/10.48550/arxiv.2108.02497.
[19] Walsh I, Pollastri G, Tosatto SCE. Correct machine learning on protein sequences: a peer-reviewing perspective. Brief Bioinform 2015;17:831–40. https://doi.org/10.1093/bib/bbv082.
[20] Greener JG, Kandathil SM, Moffat L, Jones DT. A guide to machine learning for biologists. Nat Rev Mol Cell Biol 2022;23:40–55.

https://doi.org/10.1038/s41580-021-00407-0.
[21] Sapoval N, Aghazadeh A, Nute MG, Antunes DA, Balaji A, Baraniuk R, et al. Current progress and open challenges for applying deep learning across the biosciences. Nat Commun 2022;13:1728. https://doi.org/10.1038/s41467-022-29268-7.
[22] Johnston KE, Fannjiang C, Wittmann BJ, Hie BL, Yang KK, Wu Z. Machine Learning for Protein Engineering. ArXiv 2023. https://doi.org/10.48550/arxiv.2305.16634.
[23] Xu Y, Verma D, Sheridan RP, Liaw A, Ma J, Marshall NM, et al. Deep Dive into Machine Learning Models for Protein Engineering. J Chem Inf Model 2020;60:2773–90. https://doi.org/10.1021/acs.jcim.0c00073.
[24] Kouba P, Kohout P, Haddadi F, Bushuiev A, Samusevich R, Sedlar J, et al. Machine Learning-Guided Protein Engineering. ACS Catal 2023. https://doi.org/10.1021/acscatal.3c02743.
[25] Scannell JW, Bosley J, Hickman JA, Dawson GR, Truebel H, Ferreira GS, et al. Predictive validity in drug discovery: what it is, why it matters and how to improve it. Nat Rev Drug Discov 2022;21:915–31. https://doi.org/10.1038/s41573-022-00552-x.
[26] Bergström F, Lindmark B. Accelerated drug discovery by rapid candidate drug identification. Drug Discov Today 2019;24:1237–41. https://doi.org/10.1016/j.drudis.2019.03.026.
[27] Austin M, Burschowsky D, Chan DTY, Jenkinson L, Haynes S, Diamandakis A, et al. Structural and functional characterization of C0021158, a high-affinity monoclonal antibody that inhibits Arginase 2 function via a novel non-competitive mechanism of action. MAbs 2020;12:1801230. https://doi.org/10.1080/19420862.2020.1801230.
[28] Rossant CJ, Carroll D, Huang L, Elvin J, Neal F, Walker E, et al. Phage display and hybridoma generation of antibodies to human CXCR2 yields antibodies with distinct mechanisms and epitopes. MAbs 2014;6:1425–38. https://doi.org/10.4161/mabs.34376.
[29] Furtmann N, Schneider M, Spindler N, Steinmann B, Li Z, Focken I, et al. An end-to-end automated platform process for high-throughput engineering of next-generation multi-specific antibody therapeutics. MAbs 2021;13:1955433. https://doi.org/10.1080/19420862.2021.1955433.
[30] Rodrigues T. The good, the bad, and the ugly in chemical and biological data for machine learning. Drug Discov Today: Technol 2019;32:3–8. https://doi.org/10.1016/j.ddtec.2020.07.001.
[31] Bender A, Cortes-Ciriano I. Artificial intelligence in drug discovery: what is realistic, what are illusions? Part 2: a discussion of chemical and biological data. Drug Discov Today 2021;26:1040–52. https://doi.org/10.1016/j.drudis.2020.11.037.
[32] Bellamy H, Rehim AA, Orhobor OI, King R. Batched Bayesian Optimization for Drug Design in Noisy Environments. J Chem Inf Model 2022;62:3970–81. https://doi.org/10.1021/acs.jcim.2c00602.
[33] Wilkinson MD, Dumontier M, Aalbersberg IjJ, Appleton G, Axton M, Baak A, et al. The FAIR Guiding Principles for scientific data management and stewardship. Sci Data 2016;3:160018. https://doi.org/10.1038/sdata.2016.18.
[34] Li L, Gupta E, Spaeth J, Shing L, Jaimes R, Engelhart E, et al. Machine learning

optimization of candidate antibody yields highly diverse sub-nanomolar affinity antibody libraries. Nat Commun 2023;14:3454. https://doi.org/10.1038/s41467-023-39022-2.

[35] Yang A, Jude KM, Lai B, Minot M, Kocyla AM, Glassman CR, et al. Deploying synthetic coevolution and machine learning to engineer protein-protein interactions. Science 2023;381. https://doi.org/10.1126/science.adh1720.

[36] Mason DM, Friedensohn S, Weber CR, Jordi C, Wagner B, Meng S, et al. Deep learning enables therapeutic antibody optimization in mammalian cells by deciphering high-dimensional protein sequence space. Biorxiv 2019:617860. https://doi.org/10.1101/617860.

[37] Maloney MP, Coley CW, Genheden S, Carson N, Helquist P, Norrby P-O, et al. Negative Data in Data Sets for Machine Learning Training. Org Lett 2023;25:2945–7. https://doi.org/10.1021/acs.orglett.3c01282.

[38] Zhang W, Wang H, Feng N, Li Y, Gu J, Wang Z. Developability assessment at early-stage discovery to enable development of antibody-derived therapeutics. Antib Ther 2022;6:13–29. https://doi.org/10.1093/abt/tbac029.

[39] Xiao X, Douthwaite JA, Chen Y, Kemp B, Kidd S, Percival-Alwyn J, et al. A high-throughput platform for population reformatting and mammalian expression of phage display libraries to enable functional screening as full-length IgG. MAbs 2017;9:996–1006. https://doi.org/10.1080/19420862.2017.1337617.

[40] Geiger RS, Cope D, Ip J, Lotosh M, Shah A, Weng J, et al. "Garbage in, garbage out" revisited: What do machine learning application papers report about human-labeled training data? Quant Sci Stud 2021;2:795–827. https://doi.org/10.1162/qss_a_00144.

[41] Fourches D, Muratov E, Tropsha A. Trust, But Verify: On the Importance of Chemical Structure Curation in Cheminformatics and QSAR Modeling Research. J Chem Inf Model 2010;50:1189–204. https://doi.org/10.1021/ci100176x.

[42] Fourches D, Muratov E, Tropsha A. Trust, but Verify II: A Practical Guide to Chemogenomics Data Curation. J Chem Inf Model 2016;56:1243–52. https://doi.org/10.1021/acs.jcim.6b00129.

[43] Littmann M, Selig K, Cohen-Lavi L, Frank Y, Hönigschmid P, Kataka E, et al. Validity of machine learning in biology and medicine increased through collaborations across fields of expertise. Nat Mach Intell 2020;2:18–24. https://doi.org/10.1038/s42256-019-0139-8.

[44] Jiao Y, Du P. Performance measures in evaluating machine learning based bioinformatics predictors for classifications. Quant Biol 2016;4:320–30. https://doi.org/10.1007/s40484-016-0081-2.

[45] Erickson BJ, Kitamura F. Magician's Corner: 9. Performance Metrics for Machine Learning Models. Radiol: Artif Intell 2021;3:e200126. https://doi.org/10.1148/ryai.2021200126.

[46] Vishwakarma G, Sonpal A, Hachmann J. Metrics for Benchmarking and Uncertainty Quantification: Quality, Applicability, and Best Practices for Machine Learning in Chemistry. Trends Chem 2021;3:146–56. https://doi.org/10.1016/j.trechm.2020.12.004.

[47] Söding J, Remmert M. Protein sequence comparison and fold recognition:

progress and good-practice benchmarking. Curr Opin Struct Biol 2011;21:404–11. https://doi.org/10.1016/j.sbi.2011.03.005.
[48] Won J, Baek M, Monastyrskyy B, Kryshtafovych A, Seok C. Assessment of protein model structure accuracy estimation in CASP13: Challenges in the era of deep learning. Proteins: Struct, Funct, Bioinform 2019;87:1351–60. https://doi.org/10.1002/prot.25804.
[49] Minot M, Reddy ST. Meta Learning Improves Robustness and Performance in Machine Learning-Guided Protein Engineering. BioRxiv 2023:2023.01.30.526201. https://doi.org/10.1101/2023.01.30.526201.
[50] Pavlović M, Hajj GSA, Kanduri C, Pensar J, Wood M, Sollid LM, et al. Improving generalization of machine learning-identified biomarkers with causal modeling: an investigation into immune receptor diagnostics. ArXiv Preprint ArXiv:220409291 2022.
[51] Kolmar SS, Grulke CM. The effect of noise on the predictive limit of QSAR models. J Cheminformatics 2021;13:92. https://doi.org/10.1186/s13321-021-00571-7.
[52] Li G, Zrimec J, Ji B, Geng J, Larsbrink J, Zelezniak A, et al. Performance of Regression Models as a Function of Experiment Noise. Bioinform Biol Insights 2021;15:11779322211020316. https://doi.org/10.1177/11779322211020315.
[53] Brown SP, Muchmore SW, Hajduk PJ. Healthy skepticism: assessing realistic model performance. Drug Discov Today 2009;14:420–7. https://doi.org/10.1016/j.drudis.2009.01.012.
[54] Campbell RM, Dymshitz J, Eastwood BJ, Emkey R, Greenen DP, Heerding JM, et al. Data Standardization for Results Management. 2004.
[55] Schisterman EF, Vexler A, Whitcomb BW, Liu A. The Limitations due to Exposure Detection Limits for Regression Models. Am J Epidemiology 2006;163:374–83. https://doi.org/10.1093/aje/kwj039.
[56] Lubin JH, Colt JS, Camann D, Davis S, Cerhan JR, Severson RK, et al. Epidemiologic Evaluation of Measurement Data in the Presence of Detection Limits. Environ Heal Perspect 2004;112:1691–6. https://doi.org/10.1289/ehp.7199.
[57] Anger LT, Wolf A, Schleifer K-J, Schrenk D, Rohrer SG. Generalized Workflow for Generating Highly Predictive in Silico Off-Target Activity Models. J Chem Inf Model 2014;54:2411–22. https://doi.org/10.1021/ci500342q.
[58] Tropsha A. Best Practices for QSAR Model Development, Validation, and Exploitation. Mol Inform 2010;29:476–88. https://doi.org/10.1002/minf.201000061.
[59] Young D, Martin T, Venkatapathy R, Harten P. Are the Chemical Structures in Your QSAR Correct? QSAR Comb Sci 2008;27:1337–45. https://doi.org/10.1002/qsar.200810084.
[60] OECD. Guidance Document on the Validation of (Quantitative) Structure-Activity Relationship [(Q)SAR] Models. OECD Ser Test Assess 2014. https://doi.org/10.1787/9789264085442-en.
[61] Muratov EN, Bajorath J, Sheridan RP, Tetko IV, Filimonov D, Poroikov V, et al. QSAR without borders. Chem Soc Rev 2020;49:3525–64. https://doi.org/10.1039/d0cs00098a.
[62] Apiletti D, Bruno G, Ficarra E, Baralis E. Data Cleaning and Semantic Improvement in Biological Databases. J Integr Bioinform 2006;3:219–29.

https://doi.org/10.1515/jib-2006-40.
[63] Chicco D. Ten quick tips for machine learning in computational biology. BioData Min 2017;10:35. https://doi.org/10.1186/s13040-017-0155-3.
[64] Walsh I, Fishman D, Garcia-Gasulla D, Titma T, Pollastri G, Group EMLF, et al. DOME: recommendations for supervised machine learning validation in biology. Nat Methods 2021;18:1122–7. https://doi.org/10.1038/s41592-021-01205-4.
[65] Jones DT. Setting the standards for machine learning in biology. Nat Rev Mol Cell Biol 2019;20:659–60. https://doi.org/10.1038/s41580-019-0176-5.
[66] Xu C, Jackson SA. Machine learning and complex biological data. Genome Biol 2019;20:76. https://doi.org/10.1186/s13059-019-1689-0.
[67] Shugay M, Britanova OV, Merzlyak EM, Turchaninova MA, Mamedov IZ, Tuganbaev TR, et al. Towards error-free profiling of immune repertoires. Nat Methods 2014;11:653–5. https://doi.org/10.1038/nmeth.2960.
[68] Pavlović M, Scheffer L, Motwani K, Kanduri C, Kompova R, Vazov N, et al. The immuneML ecosystem for machine learning analysis of adaptive immune receptor repertoires. Nat Mach Intell 2021;3:936–44. https://doi.org/10.1038/s42256-021-00413-z.
[69] Breden F, Prak ETL, Peters B, Rubelt F, Schramm CA, Busse CE, et al. Reproducibility and Reuse of Adaptive Immune Receptor Repertoire Data. Front Immunol 2017;8:1418. https://doi.org/10.3389/fimmu.2017.01418.
[70] Christley S, Aguiar A, Blanck G, Breden F, Bukhari SAC, Busse CE, et al. The ADC API: A Web API for the Programmatic Query of the AIRR Data Commons. Front Big Data 2020;3:22. https://doi.org/10.3389/fdata.2020.00022.
[71] Community TA, Rubelt F, Busse CE, Bukhari SAC, Bürckert J-P, Mariotti-Ferrandiz E, et al. Adaptive Immune Receptor Repertoire Community recommendations for sharing immune-repertoire sequencing data. Nat Immunol 2017;18:1274–8. https://doi.org/10.1038/ni.3873.
[72] Heiden JAV, Marquez S, Marthandan N, Bukhari SAC, Busse CE, Corrie B, et al. AIRR Community Standardized Representations for Annotated Immune Repertoires. Front Immunol 2018;9:2206. https://doi.org/10.3389/fimmu.2018.02206.
[73] Kramer C, Kalliokoski T, Gedeck P, Vulpetti A. The Experimental Uncertainty of Heterogeneous Public K i Data. J Med Chem 2012;55:5165–73. https://doi.org/10.1021/jm300131x.
[74] Kramer C, Dahl G, Tyrchan C, Ulander J. A comprehensive company database analysis of biological assay variability. Drug Discov Today 2016;21:1213–21. https://doi.org/10.1016/j.drudis.2016.03.015.
[75] Kalliokoski T, Kramer C, Vulpetti A, Gedeck P. Comparability of Mixed IC50 Data – A Statistical Analysis. PLoS ONE 2013;8:e61007. https://doi.org/10.1371/journal.pone.0061007.
[76] Aldeghi M, Graff DE, Frey N, Morrone JA, Pyzer-Knapp EO, Jordan KE, et al. Roughness of Molecular Property Landscapes and Its Impact on Modellability. J Chem Inf Model 2022;62:4660–71. https://doi.org/10.1021/acs.jcim.2c00903.
[77] Parks C, Gaieb Z, Amaro RE. An Analysis of Proteochemometric and Conformal Prediction Machine Learning Protein-Ligand Binding Affinity Models. Front Mol Biosci 2020;7:93. https://doi.org/10.3389/fmolb.2020.00093.

[78] Jain T, Sun T, Durand S, Hall A, Houston NR, Nett JH, et al. Biophysical properties of the clinical-stage antibody landscape. Proc National Acad Sci 2017;114:944–9. https://doi.org/10.1073/pnas.1616408114.
[79] Jain T, Boland T, Vásquez M. Identifying developability risks for clinical progression of antibodies using high-throughput in vitro and in silico approaches. Mabs 2023;15:2200540. https://doi.org/10.1080/19420862.2023.2200540.
[80] Wang D, Hensman J, Kutkaite G, Toh TS, Galhoz A, Team GS, et al. A statistical framework for assessing pharmacological responses and biomarkers using uncertainty estimates. ELife 2020;9:e60352. https://doi.org/10.7554/elife.60352.
[81] Fenoy E, Edera AA, Stegmayer G. Transfer learning in proteins: evaluating novel protein learned representations for bioinformatics tasks. Brief Bioinform 2022;23. https://doi.org/10.1093/bib/bbac232.
[82] Alley EC, Khimulya G, Biswas S, AlQuraishi M, Church GM. Unified rational protein engineering with sequence-based deep representation learning. Nat Methods 2019;16:1315–22. https://doi.org/10.1038/s41592-019-0598-1.
[83] Brandes N, Ofer D, Peleg Y, Rappoport N, Linial M. ProteinBERT: a universal deep-learning model of protein sequence and function. Bioinformatics 2022;38:2102–10. https://doi.org/10.1093/bioinformatics/btac020.
[84] Rives A, Meier J, Sercu T, Goyal S, Lin Z, Liu J, et al. Biological structure and function emerge from scaling unsupervised learning to 250 million protein sequences. Proc Natl Acad Sci 2021;118:e2016239118. https://doi.org/10.1073/pnas.2016239118.
[85] Wu Z, Johnston KE, Arnold FH, Yang KK. Protein sequence design with deep generative models. Curr Opin Chem Biol 2021;65:18–27. https://doi.org/10.1016/j.cbpa.2021.04.004.
[86] Choi Y. Artificial intelligence for antibody reading comprehension: AntiBERTa. Patterns 2022;3:100535. https://doi.org/10.1016/j.patter.2022.100535.
[87] Kawashima S, Ogata H, Kanehisa M. AAindex: Amino Acid Index Database. Nucleic Acids Res 1999;27:368–9. https://doi.org/10.1093/nar/27.1.368.
[88] Georgiev AG. Interpretable Numerical Descriptors of Amino Acid Space. J Comput Biol 2009;16:703–23. https://doi.org/10.1089/cmb.2008.0173.
[89] Wittmann BJ, Yue Y, Arnold FH. Informed training set design enables efficient machine learning-assisted directed protein evolution. Cell Syst 2021. https://doi.org/10.1016/j.cels.2021.07.008.
[90] Tilborg D van, Alenicheva A, Grisoni F. Exposing the Limitations of Molecular Machine Learning with Activity Cliffs. J Chem Inf Model 2022;62:5938–51. https://doi.org/10.1021/acs.jcim.2c01073.
[91] Janela T, Bajorath J. Rationalizing general limitations in assessing and comparing methods for compound potency prediction. Sci Rep 2023;13:17816. https://doi.org/10.1038/s41598-023-45086-3.
[92] Hsu C, Nisonoff H, Fannjiang C, Listgarten J. Learning protein fitness models from evolutionary and assay-labeled data. Nat Biotechnol 2022:1–9. https://doi.org/10.1038/s41587-021-01146-5.
[93] Raybould MIJ, Marks C, Krawczyk K, Taddese B, Nowak J, Lewis AP, et al. Five computational developability guidelines for therapeutic antibody profiling. P Natl

Acad Sci Usa 2019;116:4025–30. https://doi.org/10.1073/pnas.1810576116.
[94] Ahmed L, Gupta P, Martin KP, Scheer JM, Nixon AE, Kumar S. Intrinsic physicochemical profile of marketed antibody-based biotherapeutics. Proc Natl Acad Sci 2021;118:e2020577118. https://doi.org/10.1073/pnas.2020577118.
[95] Ozsahin DU, Mustapha MT, Mubarak AS, Ameen ZS, Uzun B. Impact of feature scaling on machine learning models for the diagnosis of diabetes. 2022 Int Conf Artif Intell Everything (AIE) 2022;00:87–94. https://doi.org/10.1109/aie57029.2022.00024.
[96] Wan X. Influence of feature scaling on convergence of gradient iterative algorithm. J Phys: Conf Ser 2019;1213:032021. https://doi.org/10.1088/1742-6596/1213/3/032021.
[97] Jia W, Sun M, Lian J, Hou S. Feature dimensionality reduction: a review. Complex Intell Syst 2022;8:2663–93. https://doi.org/10.1007/s40747-021-00637-x.
[98] Velliangiri S, Alagumuthukrishnan S, joseph SIT. A Review of Dimensionality Reduction Techniques for Efficient Computation. Procedia Comput Sci 2019;165:104–11. https://doi.org/10.1016/j.procs.2020.01.079.
[99] Sandve GK, Greiff V. Access to ground truth at unconstrained size makes simulated data as indispensable as experimental data for bioinformatics methods development and benchmarking. Bioinformatics 2022;38:4994–6. https://doi.org/10.1093/bioinformatics/btac612.
[100] Chen V, Yang M, Cui W, Kim JS, Talwalkar A, Ma J. Best Practices for Interpretable Machine Learning in Computational Biology. BioRxiv 2022:2022.10.28.513978. https://doi.org/10.1101/2022.10.28.513978.
[101] Marcou Q, Mora T, Walczak AM. High-throughput immune repertoire analysis with IGoR. Nat Commun 2018;9:561. https://doi.org/10.1038/s41467-018-02832-w.
[102] Weber CR, Akbar R, Yermanos A, Pavlović M, Snapkov I, Sandve GK, et al. immuneSIM: tunable multi-feature simulation of B- and T-cell receptor repertoires for immunoinformatics benchmarking. Bioinformatics 2020;36:3594–6. https://doi.org/10.1093/bioinformatics/btaa158.
[103] Robert PA, Akbar R, Frank R, Pavlović M, Widrich M, Snapkov I, et al. Unconstrained generation of synthetic antibody-antigen structures to guide machine learning methodology for real-world antibody specificity prediction. BioRxiv 2022:2021.07.06.451258. https://doi.org/10.1101/2021.07.06.451258.
[104] Jamal S, Grover A, Grover S. Machine Learning From Molecular Dynamics Trajectories to Predict Caspase-8 Inhibitors Against Alzheimer's Disease. Front Pharmacol 2019;10:780. https://doi.org/10.3389/fphar.2019.00780.
[105] Min Y, Wei Y, Wang P, Wang X, Li H, Wu N, et al. From Static to Dynamic Structures: Improving Binding Affinity Prediction with a Graph-Based Deep Learning Model. ArXiv 2022. https://doi.org/10.48550/arxiv.2208.10230.
[106] Olsen TH, Boyles F, Deane CM. Observed Antibody Space: A diverse database of cleaned, annotated, and translated unpaired and paired antibody sequences. Protein Sci 2022;31:141–6. https://doi.org/10.1002/pro.4205.
[107] Corrie BD, Marthandan N, Zimonja B, Jaglale J, Zhou Y, Barr E, et al. iReceptor: A platform for querying and analyzing antibody/B-cell and T-cell receptor repertoire data across federated repositories. Immunol Rev

2018;284:24–41. https://doi.org/10.1111/imr.12666.
[108] Janin J. Welcome to CAPRI: A Critical Assessment of PRedicted Interactions. Proteins: Struct, Funct, Bioinform 2002;47:257–257. https://doi.org/10.1002/prot.10111.
[109] Janin J. Assessing predictions of protein–protein interaction: The CAPRI experiment. Protein Sci 2005;14:278–83. https://doi.org/10.1110/ps.041081905.
[110] Kryshtafovych A, Schwede T, Topf M, Fidelis K, Moult J. Critical assessment of methods of protein structure prediction (CASP)—Round XIV. Proteins: Struct, Funct, Bioinform 2021;89:1607–17. https://doi.org/10.1002/prot.26237.
[111] Armer C, Kane H, Cortade D, Estell D, Yusuf A, Sanka R, et al. The Protein Engineering Tournament: An Open Science Benchmark for Protein Modeling and Design. ArXiv 2023.
[112] Walters WP. Chemoinformatics for Drug Discovery 2013:1–31. https://doi.org/10.1002/9781118742785.ch1.
[113] Estabrooks A, Jo T, Japkowicz N. A Multiple Resampling Method for Learning from Imbalanced Data Sets. Comput Intell 2004;20:18–36. https://doi.org/10.1111/j.0824-7935.2004.t01-1-00228.x.
[114] Cao H, Li X-L, Woon DY-K, Ng S-K. Integrated Oversampling for Imbalanced Time Series Classification. IEEE Trans Knowl Data Eng 2013;25:2809–22. https://doi.org/10.1109/tkde.2013.37.
[115] Anand A, Pugalenthi G, Fogel GB, Suganthan PN. An approach for classification of highly imbalanced data using weighting and undersampling. Amino Acids 2010;39:1385–91. https://doi.org/10.1007/s00726-010-0595-2.
[116] Esposito C, Landrum GA, Schneider N, Stiefl N, Riniker S. GHOST: Adjusting the Decision Threshold to Handle Imbalanced Data in Machine Learning. J Chem Inf Model 2021;61:2623–40. https://doi.org/10.1021/acs.jcim.1c00160.
[117] Haixiang G, Yijing L, Shang J, Mingyun G, Yuanyue H, Bing G. Learning from class-imbalanced data: Review of methods and applications. Expert Syst Appl 2017;73:220–39. https://doi.org/10.1016/j.eswa.2016.12.035.
[118] Kaur H, Pannu HS, Malhi AK. A Systematic Review on Imbalanced Data Challenges in Machine Learning. ACM Comput Surv (CSUR) 2019;52:1–36. https://doi.org/10.1145/3343440.
[119] Kumar P, Bhatnagar R, Gaur K, Bhatnagar A. Classification of Imbalanced Data:Review of Methods and Applications. IOP Conf Ser: Mater Sci Eng 2021;1099:012077. https://doi.org/10.1088/1757-899x/1099/1/012077.
[120] García V, Sánchez JS, Mollineda RA. Trends in Applied Intelligent Systems, 23rd International Conference on Industrial Engineering and Other Applications of Applied Intelligent Systems, IEA/AIE 2010, Cordoba, Spain, June 1-4, 2010, Proceedings, Part I. Lect Notes Comput Sci 2010:541–9. https://doi.org/10.1007/978-3-642-13022-9_54.
[121] Sheridan RP. The Relative Importance of Domain Applicability Metrics for Estimating Prediction Errors in QSAR Varies with Training Set Diversity. J Chem Inf Model 2015;55:1098–107. https://doi.org/10.1021/acs.jcim.5b00110.
[122] Sugita S, Ohue M. Drug-target affinity prediction using applicability domain based on data density 2021. https://doi.org/10.26434/chemrxiv-2021-hp2p9-v2.
[123] Langevin M, Grebner C, Güssregen S, Sauer S, Li Y, Matter H, et al. Impact of

Applicability Domains to Generative Artificial Intelligence. ACS Omega 2023;8:23148–67. https://doi.org/10.1021/acsomega.3c00883.

[124] Renz P, Rompaey DV, Wegner JK, Hochreiter S, Klambauer G. On failure modes in molecule generation and optimization. Drug Discov Today: Technol 2019;32:55–63. https://doi.org/10.1016/j.ddtec.2020.09.003.

[125] Langevin M, Vuilleumier R, Bianciotto M. Explaining and avoiding failure modes in goal-directed generation of small molecules. J Cheminformatics 2022;14:20. https://doi.org/10.1186/s13321-022-00601-y.

[126] Motulsky HJ, Brown RE. Detecting outliers when fitting data with nonlinear regression – a new method based on robust nonlinear regression and the false discovery rate. BMC Bioinform 2006;7:123. https://doi.org/10.1186/1471-2105-7-123.

[127] Robinson MC, Glen RC, Lee AA. Validating the validation: reanalyzing a large-scale comparison of deep learning and machine learning models for bioactivity prediction. J Comput-Aided Mol Des 2020;34:717–30. https://doi.org/10.1007/s10822-019-00274-0.

[128] López OAM, López AM, Crossa J. Multivariate Statistical Machine Learning Methods for Genomic Prediction 2022:109–39. https://doi.org/10.1007/978-3-030-89010-0_4.

[129] Hastie T, Tibshirani R, Friedman JH, Friedman JH. The elements of statistical learning: data mining, inference, and prediction. vol. 2. New York: Springer; 2009.

[130] Ozenne B, Subtil F, Maucort-Boulch D. The precision–recall curve overcame the optimism of the receiver operating characteristic curve in rare diseases. J Clin Epidemiology 2015;68:855–9. https://doi.org/10.1016/j.jclinepi.2015.02.010.

[131] Schenker N, Gentleman JF. On Judging the Significance of Differences by Examining the Overlap Between Confidence Intervals.  Am Stat 2001;55:182–6. https://doi.org/10.1198/000313001317097960.

[132] Davis J, Goadrich M. The relationship between Precision-Recall and ROC curves. Proc 23rd Int Conf Mach Learn - ICML '06 2006:233–40. https://doi.org/10.1145/1143844.1143874.

[133] Xu W, Dai J, Hung YS, Wang Q. Estimating the area under a receiver operating characteristic (ROC) curve: Parametric and nonparametric ways. Signal Process 2013;93:3111–23. https://doi.org/10.1016/j.sigpro.2013.05.010.

[134] Grandini M, Bagli E, Visani G. Metrics for Multi-Class Classification: an Overview. ArXiv 2020. https://doi.org/10.48550/arxiv.2008.05756.

[135] Sokolova M, Lapalme G. A systematic analysis of performance measures for classification tasks. Inf Process Manag 2009;45:427–37. https://doi.org/10.1016/j.ipm.2009.03.002.

[136] Delgado R, Tibau X-A. Why Cohen's Kappa should be avoided as performance measure in classification. PLoS ONE 2019;14:e0222916. https://doi.org/10.1371/journal.pone.0222916.

[137] Bishop CM, Nasrabadi NM. Pattern recognition and machine learning. vol. 4. Springer; 2006.

[138] Mullick SS, Datta S, Dhekane SG, Das S. Appropriateness of performance indices for imbalanced data classification: An analysis. Pattern Recognit 2020;102:107197. https://doi.org/10.1016/j.patcog.2020.107197.

[139] Pedregosa F, Varoquaux G, Gramfort A, Michel V, Thirion B, Grisel O, et al. Scikit-learn: Machine Learning in Python. Journal of Machine Learning Research 2011;12:2825–30.

[140] Raju VNG, Lakshmi KP, Jain VM, Kalidindi A, Padma V. Study the Influence of Normalization/Transformation process on the Accuracy of Supervised Classification. 2020 Third Int Conf Smart Syst Inven Technol (ICSSIT) 2020;00:729–35. https://doi.org/10.1109/icssit48917.2020.9214160.

[141] Amorim LBV de, Cavalcanti GDC, Cruz RMO. The choice of scaling technique matters for classification performance. Appl Soft Comput 2023;133:109924. https://doi.org/10.1016/j.asoc.2022.109924.

[142] Patro S, Sahu KK. Normalization: A preprocessing stage. ArXiv Preprint ArXiv:150306462 2015.

[143] Chuang KV, Keiser MJ. Adversarial Controls for Scientific Machine Learning. ACS Chem Biol 2018;13:2819–21. https://doi.org/10.1021/acschembio.8b00881.

[144] Rücker C, Rücker G, Meringer M. y-Randomization and Its Variants in QSPR/QSAR. J Chem Inf Model 2007;47:2345–57. https://doi.org/10.1021/ci700157b.

[145] Tropsha A, Gramatica P, Gombar VK. The Importance of Being Earnest: Validation is the Absolute Essential for Successful Application and Interpretation of QSPR Models. QSAR Comb Sci 2003;22:69–77. https://doi.org/10.1002/qsar.200390007.

[146] Lipiński PFJ, Szurmak P. SCRAMBLE'N'GAMBLE: a tool for fast and facile generation of random data for statistical evaluation of QSAR models. Chem Pap 2017;71:2217–32. https://doi.org/10.1007/s11696-017-0215-7.

[147] Chuang KV, Keiser MJ. Comment on "Predicting reaction performance in C–N cross-coupling using machine learning." Science 2018;362. https://doi.org/10.1126/science.aat8603.

[148] Lyskov S, Gray JJ. The RosettaDock server for local protein–protein docking. Nucleic Acids Res 2008;36:W233–8. https://doi.org/10.1093/nar/gkn216.

[149] Weitzner BD, Jeliazkov JR, Lyskov S, Marze N, Kuroda D, Frick R, et al. Modeling and docking of antibody structures with Rosetta. Nat Protoc 2017;12:401–16. https://doi.org/10.1038/nprot.2016.180.

[150] Pierce BG, Wiehe K, Hwang H, Kim B-H, Vreven T, Weng Z. ZDOCK server: interactive docking prediction of protein–protein complexes and symmetric multimers. Bioinformatics 2014;30:1771–3. https://doi.org/10.1093/bioinformatics/btu097.

[151] Yan Y, Tao H, He J, Huang S-Y. The HDOCK server for integrated protein–protein docking. Nat Protoc 2020;15:1829–52. https://doi.org/10.1038/s41596-020-0312-x.

[152] Desta IT, Porter KA, Xia B, Kozakov D, Vajda S. Performance and Its Limits in Rigid Body Protein-Protein Docking. Structure 2020;28:1071-1081.e3. https://doi.org/10.1016/j.str.2020.06.006.

[153] Fan W, Mencius J, Du W, Fan H, Zhu H, Wei D, et al. Online bioinformatics teaching practice: Comparison of popular docking programs using SARS-CoV-2 spike RBD–ACE2 complex as a benchmark. Biochem Mol Biol Educ 2021;49:833–40. https://doi.org/10.1002/bmb.21566.

[154] Kapoor S, Narayanan A. Leakage and the Reproducibility Crisis in ML-based Science. ArXiv 2022. https://doi.org/10.48550/arxiv.2207.07048.
[155] Bernett J, Blumenthal DB, List M. Cracking the black box of deep sequence-based protein-protein interaction prediction. BioRxiv 2023:2023.01.18.524543. https://doi.org/10.1101/2023.01.18.524543.
[156] Dietterich TG. Approximate Statistical Tests for Comparing Supervised Classification Learning Algorithms. Neural Comput 1998;10:1895–923. https://doi.org/10.1162/089976698300017197.
[157] Nadeau C, Bengio Y. Inference for the Generalization Error. Mach Learn 2003;52:239–81. https://doi.org/10.1023/a:1024068626366.
[158] Bouckaert RR, Frank E. Advances in Knowledge Discovery and Data Mining, 8th Pacific-Asia Conference, PAKDD 2004, Sydney, Australia, May 26-28, 2004. Proceedings. Lect Notes Comput Sci 2004:3–12. https://doi.org/10.1007/978-3-540-24775-3_3.
[159] Berrar D. Using p-values for the comparison of classifiers: pitfalls and alternatives. Data Min Knowl Discov 2022;36:1102–39. https://doi.org/10.1007/s10618-022-00828-1.
[160] Benavoli A, Corani G, Demšar J, Zaffalon M. Time for a Change: A Tutorial for Comparing Multiple Classifiers through Bayesian Analysis. J Mach Learn Res 2017;18:2653–2688.
[161] Demšar J. Statistical Comparisons of Classifiers over Multiple Data Sets. J Mach Learn Res 2006;7:1–30.
[162] Walters WP. Comparing classification models—a practical tutorial. J Comput-Aided Mol Des 2022;36:381–9. https://doi.org/10.1007/s10822-021-00417-2.
[163] Nicholls A. Confidence limits, error bars and method comparison in molecular modeling. Part 1: The calculation of confidence intervals. J Comput-Aided Mol Des 2014;28:887–918. https://doi.org/10.1007/s10822-014-9753-z.
[164] Nicholls A. Confidence limits, error bars and method comparison in molecular modeling. Part 2: comparing methods. J Comput-Aided Mol Des 2016;30:103–26. https://doi.org/10.1007/s10822-016-9904-5.
[165] Bender A, Glen RC. A Discussion of Measures of Enrichment in Virtual Screening: Comparing the Information Content of Descriptors with Increasing Levels of Sophistication. J Chem Inf Model 2005;45:1369–75. https://doi.org/10.1021/ci0500177.
[166] Lopes JCD, Santos FM dos, Martins-José A, Augustyns K, Winter HD. The power metric: a new statistically robust enrichment-type metric for virtual screening applications with early recovery capability. J Cheminformatics 2017;9:7. https://doi.org/10.1186/s13321-016-0189-4.
[167] Huang N, Shoichet BK, Irwin JJ. Benchmarking Sets for Molecular Docking. J Med Chem 2006;49:6789–801. https://doi.org/10.1021/jm0608356.
[168] Järvelin K, Kekäläinen J. Cumulated gain-based evaluation of IR techniques. ACM Trans Inf Syst (TOIS) 2002;20:422–46. https://doi.org/10.1145/582415.582418.
[169] Sheridan RP, Feuston BP, Maiorov VN, Kearsley SK. Similarity to Molecules in the Training Set Is a Good Discriminator for Prediction Accuracy in QSAR. J Chem Inf Comput Sci 2004;44:1912–28. https://doi.org/10.1021/ci049782w.

[170] Kearnes S. Pursuing a Prospective Perspective. Trends Chem 2021;3:77–9. https://doi.org/10.1016/j.trechm.2020.10.012.
[171] Sheridan RP. Time-Split Cross-Validation as a Method for Estimating the Goodness of Prospective Prediction. J Chem Inf Model 2013;53:783–90. https://doi.org/10.1021/ci400084k.
[172] Landrum GA, Beckers M, Lanini J, Schneider N, Stiefl N, Riniker S. SIMPD: an Algorithm for Generating Simulated Time Splits for Validating Machine Learning Approaches 2023. https://doi.org/10.26434/chemrxiv-2023-x9pjf.
[173] Chothia C, Lesk AM. The relation between the divergence of sequence and structure in proteins. EMBO J 1986;5:823–6. https://doi.org/10.1002/j.1460-2075.1986.tb04288.x.
[174] Li Y, Yang J. Structural and Sequence Similarity Makes a Significant Impact on Machine-Learning-Based Scoring Functions for Protein–Ligand Interactions. J Chem Inf Model 2017;57:1007–12. https://doi.org/10.1021/acs.jcim.7b00049.
[175] Li W, Godzik A. Cd-hit: a fast program for clustering and comparing large sets of protein or nucleotide sequences. Bioinformatics 2006;22:1658–9. https://doi.org/10.1093/bioinformatics/btl158.
[176] Altschul SF, Madden TL, Schäffer AA, Zhang J, Zhang Z, Miller W, et al. Gapped BLAST and PSI-BLAST: a new generation of protein database search programs. Nucleic Acids Res 1997;25:3389–402. https://doi.org/10.1093/nar/25.17.3389.
[177] Sirocco F, Tosatto SCE. TESE: generating specific protein structure test set ensembles. Bioinformatics 2008;24:2632–3. https://doi.org/10.1093/bioinformatics/btn488.
[178] Finn RD, Bateman A, Clements J, Coggill P, Eberhardt RY, Eddy SR, et al. Pfam: the protein families database. Nucleic Acids Res 2014;42:D222–30. https://doi.org/10.1093/nar/gkt1223.
[179] Nayak SK, Ojha AC. Machine Learning and Information Processing, Proceedings of ICMLIP 2019. Adv Intell Syst Comput 2020:203–12. https://doi.org/10.1007/978-981-15-1884-3_19.
[180] Krützfeldt L-M, Schubach M, Kircher M. The impact of different negative training data on regulatory sequence predictions. PLoS ONE 2020;15:e0237412. https://doi.org/10.1371/journal.pone.0237412.
[181] Schneider C, Buchanan A, Taddese B, Deane CM. DLAB—Deep learning methods for structure-based virtual screening of antibodies. Bioinformatics 2021;38:btab660-. https://doi.org/10.1093/bioinformatics/btab660.
[182] Dens C, Laukens K, Bittremieux W, Meysman P. The pitfalls of negative data bias for the T-cell epitope specificity challenge. BioRxiv 2023:2023.04.06.535863. https://doi.org/10.1101/2023.04.06.535863.
[183] Gao Y, Gao Y, Dong K, Wu S, Liu Q. Reply to: The pitfalls of negative data bias for the T-cell epitope specificity challenge. BioRxiv 2023:2023.04.07.535967. https://doi.org/10.1101/2023.04.07.535967.
[184] Montemurro A, Jessen LE, Nielsen M. NetTCR-2.1: Lessons and guidance on how to develop models for TCR specificity predictions. Front Immunol 2022;13:1055151. https://doi.org/10.3389/fimmu.2022.1055151.

[185] Grazioli F, Mösch A, Machart P, Li K, Alqassem I, O'Donnell TJ, et al. On TCR binding predictors failing to generalize to unseen peptides. Front Immunol 2022;13:1014256. https://doi.org/10.3389/fimmu.2022.1014256.

[186] Lundberg S, Lee S-I. A Unified Approach to Interpreting Model Predictions. ArXiv 2017. https://doi.org/10.48550/arxiv.1705.07874.